\documentclass[]{article}	

\usepackage[utf8]{inputenc}				
\usepackage[big,online]{dgruyter}	
\usepackage{lmodern} 
\usepackage{microtype}
\usepackage{bm}
\usepackage{subcaption}
\usepackage{makecell}

\usepackage[
  backend=biber,      
  style=authoryear,   
  natbib=true,        
  uniquename=false,   
  maxcitenames=2      
]{biblatex}
\usepackage[table]{xcolor}
\usepackage{soul}

\definecolor{red}{rgb}{0,0,0} 

\theoremstyle{dgthm}

\theoremstyle{dgdef}

\begin{document}


\title{Investigating Experiential 
Effects in Online Chess using a Hierarchical Bayesian Analysis}
\runningtitle{Experiential Effects in Online Chess}

\author[1]{Adam Gee}
\author[2]{Sydney O. Seese}
\author[2]{James P. Curley}
\author[1]{Owen G. Ward}
\runningauthor{Gee et al.}
\affil[1]{\protect\raggedright 
Department of Statistics and Actuarial Science, Simon Fraser University, Burnaby, Canada e-mail: adam\_gee@sfu.ca,
owen\_ward@sfu.ca}
\affil[2]{\protect\raggedright 
Department of Psychology, University of Texas at Austin, Austin, USA, e-mail: sydney.seese@utexas.edu, curley@utexas.edu}
	
	
\abstract{The presence or absence of winner-loser effects is a widely 
discussed phenomenon across both sports and
psychology research. Investigation of such effects is often hampered by the 
limited availability of data.
Online chess has exploded in popularity in recent years and provides vast 
amounts of data which can be 
used to explore this
question. With a hierarchical Bayesian regression model,
we carefully investigate the 
presence of such experiential effects in online
chess. Using a large quantity of online chess data, we see little evidence 
for experiential effects that are consistent
across all players, with some individual players showing some evidence for such 
effects. Given the challenging temporal nature of this data, we discuss several
methods for assessing the suitability of our model and
carefully check its validity.}


\maketitle

\section{Introduction} 
It is rare that the success or failure of an individual competitor or sports team 
is not discussed in the context of their recent performance. The idea
of momentum in competition,
broadly defined, has been examined repeatedly in the scientific 
literature. In sports analytics this is often investigated in the context of
the ``hot hand'' phenomenon, where players experience streaks of 
success that appear
unlikely to occur due to chance \citep{miller2018surprised,otting2020hot, pelechrinis2022hot}.
Similarly, other studies have directly considered the question of momentum 
in sport, both within and between games 
\citep{gauriot2018psychological,steeger2021winning}.

A large body of research in the psychology and animal behavior
literature  explores a specific formulation of this concept of
momentum. It is commonly referred to
as  winner-loser or experiential  
effects\footnote{We will use these two names interchangeably throughout this work.},
defined as success (or failure) in recent competition
leading to an increased  probability of success (or failure) in future competition \citep{dugatkin1997winner}. 
The presence of these winner-loser effects have been
hypothesized to facilitate the formation of 
stable dominance hierarchies \citep{dugatkin1997winner, kura2015modelling}.
In animals, loser effects appear to consistently be stronger than winner effects 
(e.g. sticklebacks \citep{bakker1986aggressiveness};
crayfish \citep{bergman2003field};
copperhead snakes \citep{schuett1997body}).
In humans, several studies have revealed some evidence for  
winner and loser effects in competitive sports including
tennis, football and judo 
\citep{cohen2017choking, gauriot2018psychological, gauriot2019does, page2017winner}.
For example, winners of tennis tie-breaks are more likely to win subsequent sets when players
are evenly matched \citep{page2017winner}. Winner and loser effects have also been experimentally
demonstrated in humans playing competitive video games \citep{smith2024winner}. 
However, there is still considerable discussion 
about the existence and consistency of these effects across individuals, contexts and species. 

Indeed, there are still several outstanding key questions regarding the presence of such experiential effects, including
\begin{itemize}
    \item Are winner/loser effects strongest when considering the immediate previous outcome or several previous outcomes? 
    \item Do the size of winner/loser effects vary across rating levels such as between players/teams of stronger versus lesser ability?
    \item Are winner/loser effects stable across time?
    \item Do winner/loser effects  vary  within individuals across contexts such as across different formats of competitions?
    
\end{itemize}

One of the main challenges posed by these questions is the availability of appropriate 
data to investigate these phenomena.
Online chess has enjoyed widespread popularity in recent years,
with millions of regularly active users on several major sites.
The most common format of chess played on these
sites is ``Standard'' rated games. 
These games are often short ``bullet'' games, lasting less than three 
minutes. It is common for players to play many such games in succession.
In these games, players
are randomly
paired with another user currently wishing to play at the same time and 
of similar estimated ability. 
The color that each player is assigned (White or Black) is randomly chosen.
Games of this form
provide an excellent setting to investigate the presence of psychological momentum 
and experiential effects in such sports 
and can be used to consider the key questions posed above.
Any individual plays a wide range of players, who may utilise different 
styles and strategies of chess.
The random assignment of opponents and individual game settings, coupled 
with the vast amount of games 
many users play, makes online chess an excellent setting to 
investigate the presence of winner-loser effects 
in a controlled environment.

In this paper we use a Bayesian hierarchical logistic regression model to 
investigate the presence
of winner-loser effects in online chess, using publicly available data. 
We carefully motivate and describe the model chosen.
Then, we use this model to 
examine if and when 
these effects occur, and how they vary across player ability and game format.
While we do not believe this model is a complete representation
of the process underlying these games, we use multiple model 
checking procedures to ensure our model is reasonable.

We find little evidence for strong persistent winner-loser effects,
with this conclusion consistent across the range of players and abilities 
analysed. Individuals appear to show some variability, 
and there is some evidence that some players demonstrate such effects. 
While there is somewhat increased evidence for these winner-loser effects
in the top international competitors,
these
effects do not appear to be systematic
across the population of chess players analysed.

This article is organized as follows. We first provide a brief history 
of the study of these effects across disciplines in Section~\ref{sec:hist}.
In Section~\ref{sec:data} we describe in detail the data we use here,
highlighting key properties which 
make it well suited for the study of winner-loser effects.
Section~\ref{sec:model} gives the proposed hierarchical Bayesian regression model,
detailing the model and prior specification.
In Section~\ref{sec:fitted} we fit the proposed model, highlighting 
the results for two cohorts of chess players of differing abilities.
Section~\ref{sec:checking} provides a detailed discussion 
of  one natural procedure used to ensure our model is 
able to correctly capture the temporal structure present 
in such data.  We include several
further model checking approaches in the appendix. Finally, in Section~\ref{sec:summary}
we summarise the results of this model and highlight important
considerations for future work.


\subsection{A history of the study of experiential effects in human and animal competition}
\label{sec:hist}

Winner-loser effects have consistently been demonstrated across species engaged in various competitive interactions \citep{trannoy2017fruitfly, schwartzerhamsters2013, lanhsu2011, oldhampigs2020, hsuwolf1999, lerenarank2021, oliveirandrogen2009, abecrickets2021, kasumovic2010, oliveirazebrafish2011, bakkerstickleback1983, dugatkinwi2004, stevensonterritory2013, lehnerrats2011}.
The most common experimental design utilizes a series of dyadic matchups whereby individuals are induced to win 
or lose against weaker or stronger opponents and are then evaluated in subsequent competitions. 
These studies have demonstrated that loser effects are typically more long lasting than winner effects 
\citep{trannoy2017fruitfly, lanhsu2011, abecrickets2021, stevensonterritory2013, lehnerrats2011}. 
The likelihood of winning subsequent encounters has also been shown to increase following unresolved aggressive 
encounters (e.g. East African cichlid fish \citep{dijkstracichlid2012}) or even following recent losses 
(olive fruit fly \citep{benellifly2015}). Further, variation in competitive contexts or individual variability 
in skill level of fighting \citep{briffaskill2017}, growth rate \citep{limangrovefish2023}, age 
\citep{fawcettexp2010}, and life experience with fighting \citep{laskowskifish2016} can all influence the 
strength of winner-loser effects. Functionally, loser effects may serve to reduce aggression intensity within 
groups, with fewer attacks received by previous losers, resulting in fewer injuries to losers 
\citep{stevensonterritory2013, lehnerrats2011}. Winner effects serve to protect dominants from excessive energy 
usage by resolving aggressive encounters more efficiently \citep{lehnerrats2011}. Several theoretical models 
have demonstrated that winner-loser effects are sufficient to lead to the formation of stable dominance 
hierarchies \citep{dugatkin1997winner, dugatkindom2004}.

There exists relatively less evidence for the presence of winner-loser effects in human 
competitive interactions despite their existence being strongly hypothesized. 
In humans, these effects are also referred to as psychological momentum which is defined as a change in the 
probability of 
an outcome occurring as a function of the outcome of preceding events \citep{adlermoment1981}. For example, it 
has been 
reported that male judoists are more likely to win their next contest if they won their previous event 
\citep{cohenzada2017}, and male tennis players are more likely to win the next point if they won the preceding 
point 
\citep{gauriot2019does}. Male tennis players are also more likely to win a subsequent set against an evenly 
matched 
opponent after winning a close tiebreak \citep{page2017winner}. These effects were not observed in female 
players. Similar 
instances of psychological momentum have been observed from large datasets of competitive sports, including that 
hitting 
the bullseye in elite archery improves performance on the next shot \citep{zhaoarchery2023}, and that the 
likelihood of 
winning double-headers in softball and baseball increases if both games are played on the same day rather than 
over 
multiple days \citep{eldakarbaseball2022, gallupbaseball2018}. Alternatively, researchers have experimentally manipulated 
the outcomes of competitive interactions in video games and shown that individuals who are randomly assigned to 
be winners 
perform better in future contests than do those who are randomly assigned to be losers \citep{smith2024winner}. 


Notably, there has been a relative lack of investigation of human competitive interactions that incorporate 
thousands of competitive interactions over a longer 
period of time such as is available with data from online chess matches.
Similarly, the existing models for examining such effects are often limited or somewhat
domain specific. 
{\color{red} While models such as \citet{lerenarank2021} examine 
experiential effects in animals, their approach is largely exploratory, 
examining the relationship between several potential outcomes of interest and previous 
performance using correlation and generalized linear mixed models. This 
form of approach, along with traditional statistical tests,
is common in the animal literature. In sports, models which
have been proposed cannot easily be modified to other scenarios. While an 
advanced statistical model was proposed to investigate the hot hand in professional
darts \citep{otting2020hot}, this utilises the fact that each player has three throws
in each turn, examining experiential effects both within a turn and across a leg (several turns).
It is unclear how to extend this to more general forms of competition or interactions.
Similarly, \citet{gauriot2018psychological} use causal inference techniques
to examine the impact of teams scoring before half time in soccer, matching
these goals to similar attempts which instead hit the post. However, it may be challenging
to modify this matching approach to other games or scenarios.}
In this work we develop a natural statistical model for such 
interaction data which can be applied to any dataset of repeated competition. 
As discussed below, it can also incorporate information about these competitions as available.


\section{Data}
\label{sec:data}

\begin{figure}[ht]
    \centering
    \includegraphics[width=0.75\textwidth]{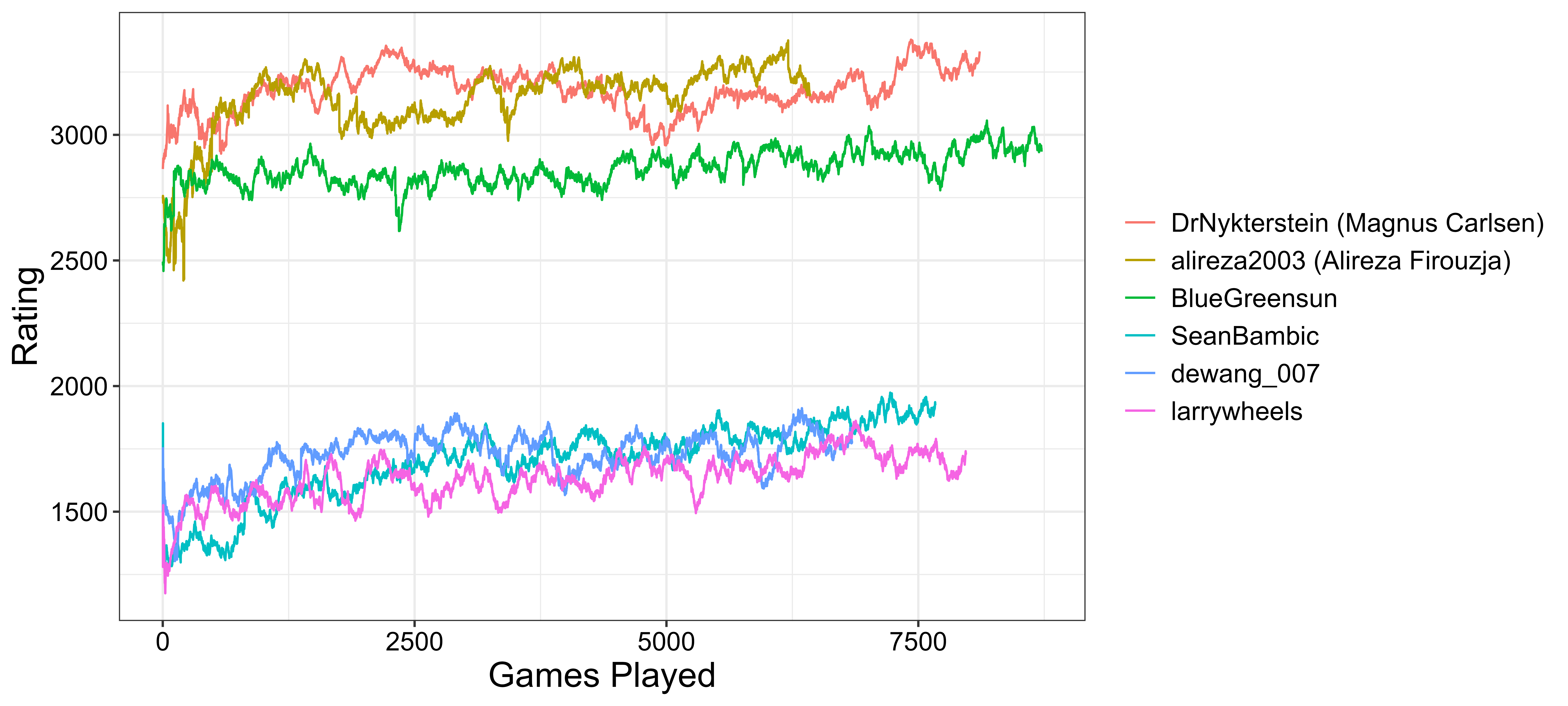}
    \caption{Evolution of Bullet Glicko-2 ratings for 3 players in the GM cohort and 3 players in the 1700-1900 cohort. ``DrNykterstein" (Magnus Carlsen) and ``alireza2003" (Alireza Firouzja) are among the current top players in the world, with Carlsen being at the number 1 spot since 2011.}
    \label{fig:elochange}
\end{figure}

In this work we use rated standard games played on Lichess, a popular open source chess server. Lichess and other chess platforms pair random opponents
who are believed to have a similar chess ability, 
to create games where both players have an opportunity to win. In online and over the board chess, it is common to estimate
the ability of an individual using a numeric rating based on their 
past performance.
When an individual creates an account on Lichess,
they are a given a default rating of 
1500 across all chess formats.
As the player plays rated games at a specific time format, their rating will update 
based on their
performance and the rating of their opponents. 
Lichess uses the Glicko-2 rating system 
{\color{red}\citep{glickman2001dynamic, glickman2024models}}, which can also estimate a measure of variability of this rating. 
{\color{red} This new system builds on the original system of \citet{glickman1999parameter} by producing larger variability in the estimated ratings if unexpected results occur.}
In Figure~\ref{fig:elochange} we show the rating evolution of six players with a similar number of games played from the two cohorts we analyse in detail below.
We include three players who are among the highest rated on Lichess (two
of whom are known elite players). We also include three players of a more moderate
ability. {\color{red}These players show some of 
the typical behaviour across the range of
players we consider. These ratings can show large changes initially, with smaller
changes present after several hundred games are played. Similarly, 
some of these players, such as ``SeanBambic", can show a gradual upward trend,
indicating gradual improvement in skill over time.}


Lichess allows a wide range of variants and time formats of chess, with
games ranging from several minutes
to several days. The only variants we consider are standard chess formats,
which is the vast majority of games on Lichess.
In these time controlled chess games, each player starts with the same 
amount of time to make their moves. 
The two most common formats are ``Bullet'' 
(games where each player has 1-2 minutes) and ``Blitz'' 
(games where each player has 3-5 minutes). 
These short games commonly end with one player winning via 
checkmate, resignation or running
out of time. Draws are possible but relatively rare in these formats. 
 We consider only the most played time controls in bullet and blitz for analysis: 60+0 
and 180+0. The first number in these time controls represents 
how many seconds each player starts with to make all their moves. The second represents 
the increment, which is the number of additional seconds added to their remaining time after making each move.

We wish to examine the presence of experiential effects across a range of 
chess abilities 
and playing history. Using a players current rating, we collected data for 40 players in each of 3 separate rating 
cohorts: 1700-1900, 2000-2200 and 2300-2500. Between October 23-31, 2023, 
we randomly selected 40
players within each of these rating ranges that had recently played 
bullet games of any format,
and scraped information about the games they had played up to that date. These rating cohorts correspond
to a range of moderate to high chess ability\footnote{Lichess provides a weekly distribution of ratings
on the platform. At the time of writing, 47\% of bullet players had a rating lower than 1500, while 
only 1.5\% had a rating higher than 2500.}. 
We also collect all games played by 25 grandmaster (GM) accounts on Lichess in early 2024.
Grandmaster is an official title awarded to players who demonstrate excellent ability in chess. At the time of writing, this extremely selective title has only been awarded to 2087 players in history. 
For the GMs we selected, the average rating was 3055 for Bullet and 2854 for Blitz games. We denote the 40 players in each rating cohort and the 25 GMs
as ``focal'' players in this paper.
We have made this data, along with the code used to scrape it, publicly available in our associated Github repository.

For every game by each focal player in the 60+0 and 180+0 time controls,
we observe their current rating, the rating
of their opponent, the colour of each player and the outcome of the game. 
{\color{red}Importantly, we note that we are unable to obtain the uncertainty in a 
player's rating (known as the rating deviation),
which is a key component in the Glicko-2 rating system.}
For brevity, we focus our findings on the 
1700-1900 and GM cohorts in what follows. 
The results obtained for the
other cohorts, unless noted, agree with those presented below,
with some of the corresponding analyses
included in the appendix.

\subsection{Data Summary}

We summarize the total number of games we use for both the 1700-1900 and 
GM cohort in Table~\ref{tab:summary_games}. For the 1700-1900
cohort we have 36 focal players playing 60+0 Bullet Games. 
This is more than 87\% of all rated bullet games they play (across
all possible time limits and increments), with each averaging over 16000
games. For the grandmasters 60+0 games correspond to 75\% 
of all bullet games played, averaging over 8000 per player.
The corresponding statistics for 180+0 Blitz games are shown in Table~\ref{tab:summary_games}. Bullet games are considerably more popular across all cohorts, perhaps due to the short time required.


\begin{table}[ht]
\centering
\begin{tabular}{llcccc}
\hline
\textbf{Game Type} & \textbf{Cohort} & \textbf{{\color{red}Num. Players}} & \textbf{Games (count)} & \textbf{Percent of Games} & \textbf{Avg Games per Player} \\
\hline
Bullet (60+0) & 1700--1900  & {\color{red}36} & 590,253  & 87\%  & 16,396 \\
Bullet (60+0) & Grandmasters & {\color{red}23} & 196,522  & 75\%  & 8,544 \\
Blitz (180+0) & 1700--1900  & {\color{red}36} & 107,670  & 58\%  & 4,141 \\
Blitz (180+0) & Grandmasters & {\color{red}23} & 34,251   & 98\%  & 1,489 \\
\hline
\end{tabular}
\caption{Summary statistics for bullet and blitz games by cohort and format. Here percent of Games
corresponds to how many of all Bullet/Blitz games were of this specific time format. }
\label{tab:summary_games}
\end{table}


\subsubsection*{The time scale of experiential effects}
A key question in the study of experiential effects is if and how these effects
persist over time. In the context of teams sports such as football, there
may be several days between subsequent games, which could interfere with such 
effects. 
Due to the short time format most popular in online chess, it is common for 
players to play many games in quick succession. 
For bullet and blitz games where the players do not receive an increment,
the maximum game length is simply the sum of the time both players start with.

\begin{figure}[ht]
	\centering
    \begin{subfigure}{0.45\textwidth}
    \centering
    \includegraphics[width=\textwidth]{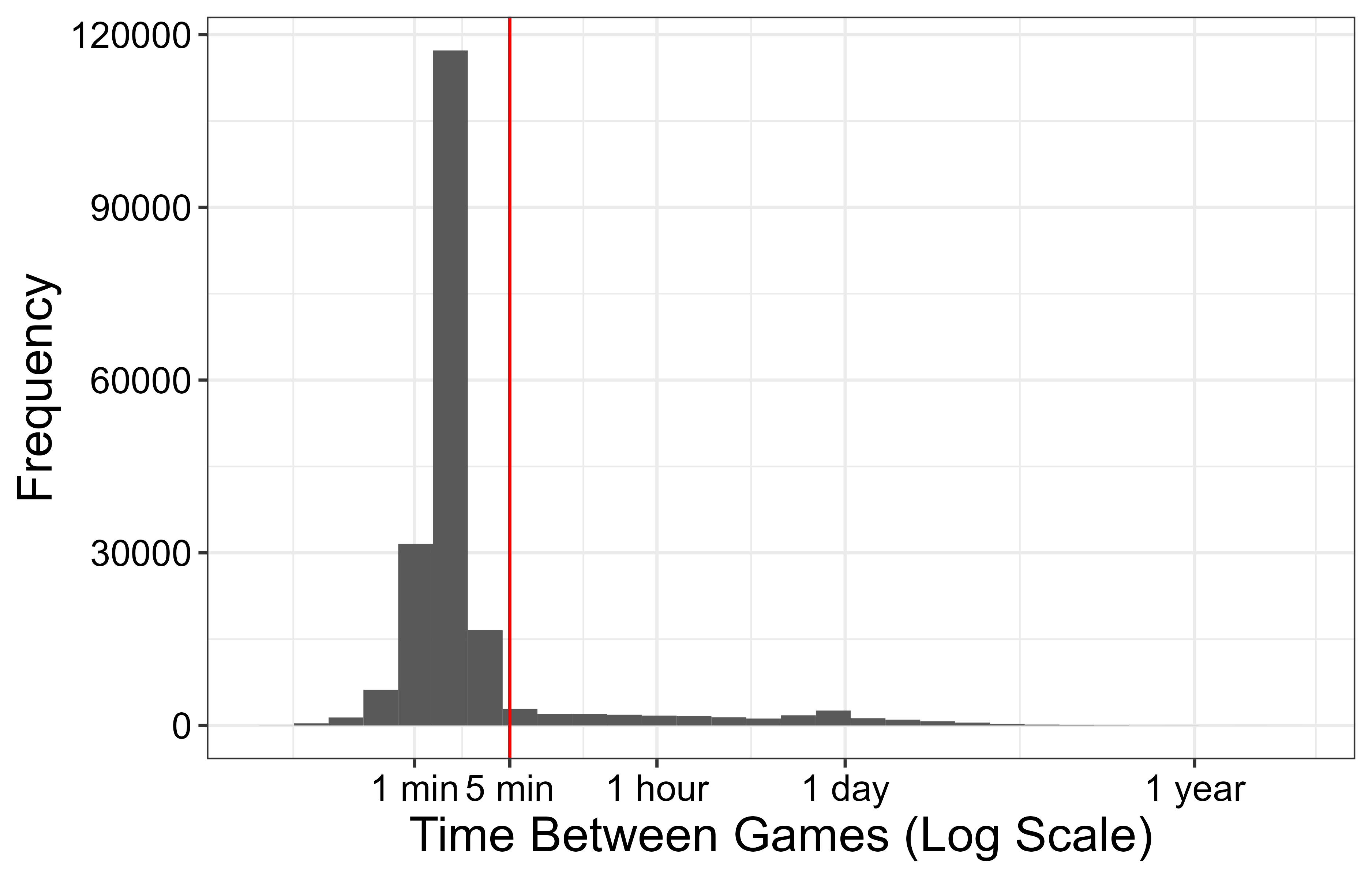}
    \end{subfigure}
    \begin{subfigure}{0.45\textwidth}
    \centering
    \includegraphics[width=\textwidth]{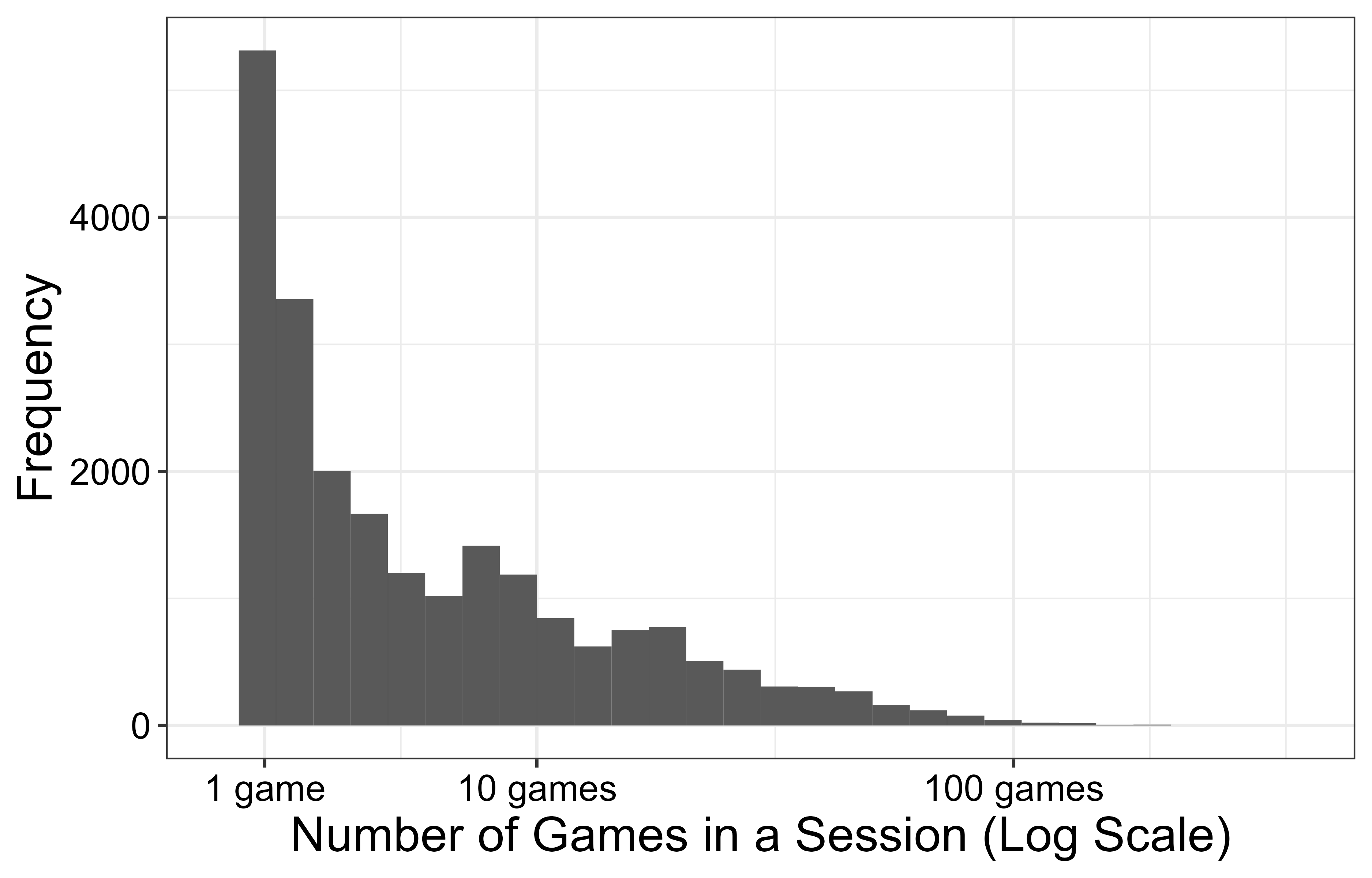}
    \end{subfigure}
    \caption{Time gaps between successive bullet games (left) and the distribution of the lengths of a session  of bullet games (right) for the GM cohort.}
	\label{fig:time_diff_session_length}
\end{figure}

In 
Figure~\ref{fig:time_diff_session_length}(left)
we show the time between the start 
of successive bullet games in the GM cohort on a log scale. 
The maximum 
length of these games is 2 minutes.
Across all grandmaster games, 88.6\%
of the games are played within 5 minutes
of the previous game.
This indicates that periods where players play more than one game 
in a row are very common (for the 1700-1900 cohort this is 85\% of 
all bullet games).

Based on this information we define bullet games played in the same session as those
which start within 5 minutes 
of the start of the previous game. With this 
definition we can then determine how many games are played in a given 
session. We plot the distribution of the number of games in a session,
from the GM cohort, in Figure~\ref{fig:time_diff_session_length}(right).
The median 
number of games played in such a session is 4, with 76.3\% of all
sessions involving more than 1 game.
It is very common for an individual to play
multiple games in quick succession. 
It is commonly thought that experiential effects
are short term, with the result of
the previous competition having an impact on
the next \citep{franz2015self}.
If experiential effects do occur, it seems plausible 
that they would be detected in games of this form,
where players who win or lose can almost immediately 
enter another contest.

While it is believed that winner-loser effects occur based on
the most recent history, we wish to 
investigate if longer term effects can exist also.
In the model described below we will investigate the
presence of these effects 
based on the result immediately proceeding 
the current game only. However, we will also 
investigate if these effects have a longer 
duration by using the performance
in the previous 10 games 
in a session
as a predictor of the current outcome.
It is still relatively common for players to play 10 or more games 
in succession, with 26.3\% of sessions played by GMs, 
as we have defined them, 
containing more than 10 games\footnote{We see similar proportions for the 1700-1900 cohort.}.
Fitting our proposed model taking account of this longer history allows investigation 
of more prolonged experiential effects.





\section{Modeling Winner-Loser Effects with a Hierarchical Logistic Regression}
\label{sec:model}

We have large volumes
of games where players are randomly assigned to opponents of similar ability
and we observe potential confounding factors such as this difference in 
ability and the colour each player plays. This makes this data
particularly well suited to investigate the role of experiential 
effects on game results.

To investigate these effects,
we model the outcome of each game played by the focal players in
our data
using a Bayesian hierarchical logistic
regression model. 
The model we consider is general and can readily be applied to many forms of data
consisting of
repeated competition over time,
identifying both individual and population level effects in such data.
The Bayesian nature of our model provides 
natural uncertainty quantification in our 
estimates along with tools for checking the 
validity of our model.

The probability a focal player wins 
an individual game is given by

\begin{equation}
P(y_{ij} = 1) = p_{ij}= \frac{1}{1 + \exp(-(\alpha_j + \beta_j \tilde{x}_{ij}^{n} +
 \bm{\gamma}^T \bm{z_{ij}}))}.
\end{equation}

Here $p_{ij}$ is the probability of the $j$-th focal player 
winning their $i$-th game, $y_{ij}$.
$\alpha_j$ is a player level random effect, indicating a player's individual ability to win a
game against an equally rated opponent when playing with the black pieces.
$\beta_j$ is a player level random effect, accounting for the win ratio $\tilde{x}_{ij}^{n}$
of the focal player over their previous $n$ games. 
We discuss this in more detail below.
$\bm{\gamma}$ is a vector of fixed effects corresponding to two individual game
covariates,
the colour played by the focal player and the difference in rating between
the focal player and their 
opponent. We partially pool the $\alpha$ and $\beta$ coefficients with a 
bivariate normal prior distribution
and place independent normal priors on the fixed effects.
The prior specification for all parameters is given by
\begin{equation}
    \begin{pmatrix}
        \alpha_j \\
        \beta_j
    \end{pmatrix}
    \sim \mathcal{N}(\bm{\mu}, \Sigma), 
\end{equation}
and 
$$
\gamma_1 \sim \mathcal{N}(0, \sigma_{g_1}^2), \ \gamma_2 \sim \mathcal{N}(0, \sigma_{g_2}^2)
$$
where we have
\begin{equation}
    \bm{\mu} = \begin{pmatrix}
        0 \\
        \mu_{\beta}
    \end{pmatrix}
    \mbox{ and }
    \Sigma = \begin{pmatrix}
        \tau_1 & 0 \\
        0  & \tau_2
    \end{pmatrix}
    \Omega 
    \begin{pmatrix}
        \tau_1 & 0 \\
        0  & \tau_2
    \end{pmatrix}.
\end{equation}

We then place independent hyperpriors on each of these parameters, with
$$
\mu_{\beta} \sim \mathcal{N}(0, \sigma^2),
$$

$$
\tau_1,\tau_2, \sigma, \sigma_{g_1}, \sigma_{g_2} \sim \mathcal{N}_{+}(0, 1)
$$
and 
$$
\Omega \sim \mbox{LKJcorr}(2),
$$

{

{\color{red} where $\mbox{LKJcorr}(2)$ is a Lewandowski-Kurowicka-Joe (LKJ) distribution with shape parameter 2 \citep{lewandowski2009generating} and $\mathcal{N}_{+}(0, 1)$ is a positively truncated standard normal distribution.}

As we are interested in identifying potential global winner-loser effects, 
we place a prior
on $\mu_{\beta}$, the global winner-loser effect.  
The aim of inferring this parameter is to 
identify if there is evidence 
for a common effect across all players, 
rather than isolated to individual players.
$\mu_{\beta} > 0$ would indicate the presence of a cohort level winner-loser effect.
If the focal player had won (lost) more recent games than their overall win
proportion, this would lead to their probability of winning 
the next game increasing (decreasing). $\mu_{\beta} < 0$ 
would indicate a reverse relationship, where poor recent 
performance would lead to an increased win probability 
in future games. 
We note that this model gives equal weight to both these winner and loser effects
and cannot identify them separately, an important natural extension.
This model can then be fit for various choices of $n$, the number of previous 
games considered in $\tilde{x}_{ij}^{n}$.
As discussed above, we will consider $n=1$ and $n=10$, investigating both
immediate and longer term
experiential effects. In the appendix we demonstrate 
similar results with $n=5$ also.


We normalize the win ratio by the cumulative win ratio of player $j$ up 
to the the $i$-th game.
If $x_{ij}^n$ is the proportion of games the $j$-th player has won in the past $n$ games then $\tilde{x}_{ij}^n = x_{ij}^n - \bar{x}_j$, where $\bar{x}_j$ is their cumulative win proportion up to the current game.
Importantly,
we consider previous games to only be those played in 
the current session, as defined previously. If this is the first game in
a session then we define $\tilde{x}_{ij}^{n}=0$. Note that if 
$n=1$, meaning we only consider the previous game in the session, then
$\tilde{x}_{ij}^{n}$ will be positive if the previous game was a victory,
and negative if the previous game was a loss. 

For $\bm{z_{ij}}$ we incorporate 2 covariates, namely the colour played
by the focal player in an individual game and the rating difference between the focal 
player and their opponent. 
$\bm{\gamma} = (\gamma_1, \gamma_2)$ where
$\gamma_1$ describes how the win probability changes in going
from playing as black to playing as white. $\gamma_1 > 0$ would indicate
that against an equally rated opponent, there would be an increased probability of winning 
when playing as white compared to playing as black.
$\gamma_2$ describes the relationship between 
win probability and the 
difference in rating between
the focal player and their current opponent. 
$\gamma_2>0$ would indicate that the more highly rated a focal player is than
their
opponent, the higher the probability they would win a match, 
accounting for the colour assigned to each player.

We decompose the covariance matrix $\Sigma$ into a correlation matrix 
$\Omega$ and the individual scale of the two random effects. 
For the correlation matrix we choose a Lewandowski-Kurowicka-Joe (LKJ) distribution as
the prior, with shape parameter 2
\citep{lewandowski2009generating}.
This is a prior over the set of correlation matrices with the chosen shape parameter favoring less correlation.
We use independent 
 half standard normal priors for the scale
parameters $\tau_1,\tau_2$.
In initial model fitting we found that less informative priors for the scale parameters,
 such as Inverse-Gamma,
gave equivalent results while making posterior convergence more challenging.


This model is a natural approach to investigating this phenomenon in online chess, and builds on models which have been considered in both the sport analytics and animal behaviour literature. 
{\color{red}
\citet{franz2015self} examine experiential effects in primates by modifying the ELO rating
    system. They modify the K-factor in this model, estimating this
    as a function of contest level predictors. This model could be applied to
    similar scenarios to the one we consider, however the ELO rating has largely
    been superseded by the Glicko-2 method. Our Bayesian approach also
    provides additional uncertainty quantification for identifying experiential effects.
    The model for hockey goalie performance of \citet{ding_hockey} is similar to our
    approach, in that they fit a hierarchical logistic regression to contest
    outcomes and include historic performance as a predictor. Our model expands on 
    this approach by considered a more flexible modeling structure, 
    specifically designed to identify a global experiential effect. We
    also develop novel model checking methods.
}
The flexibility of our model also 
allows it to be applied to a range of potential datasets consisting of repeated competition observed for a 
number of individuals.


\section{The Fitted Model}
\label{sec:fitted}

We fit the proposed Bayesian model with Stan \citep{stan2024},
using the CmdStanR \citep{cmdstanr2024} interface in R \citep{r2024}.
We use default settings for fitting these models, running 4 chains each with 1000 warmup 
and 1000 sampling iterations. 
We fit the proposed model for each of the four cohorts discussed previously.
In each case the models indicate posterior convergence, as indicated by $\hat{R}$ values close 
to 1 and large effective sample sizes \citep{gelman2013bayesian}.
We discuss the results of these models for two cohorts: the 1700-1900 cohort and the GM cohort.
Similar results for the other cohorts are included in the appendix.
All code required to reproduce this analysis will be available in the associated Github
repository, along with access to the data used.

\paragraph*{Comparing the fitted model for players of different ability.}
We first fit the proposed model for standard bullet games
as described above with $n=1$,
utilising the
previous game in the same session as the only history.
We examine the key global parameters ($\gamma_1,\gamma_2,\mu_{\beta}$),
in all four cohorts, with posterior credible intervals 
for each shown in Figure~\ref{17_19_gm_global}.
The main difference is seen in 
the fixed effects $\gamma_1$ and $\gamma_2$.
The estimated posterior mean of the colour effect ($\gamma_1$) is 
0.087 and 0.23 for the 1700-1900 and GM cohorts respectively. For a player with a win probability of 0.5 playing as black against an equally rated opponent, playing as white would increase the win probability on average to 0.56 for the GMs but only 0.52 for the 1700-1900s. This agrees with common beliefs for the first move advantage for GMs. Among decisive results in top GM games, white wins 54-56\% of games,
compared to black winning 44-46\% of games. It is also commonly believed that this advantage diminishes at lower levels of play {\color{red}\citep{kaufman2021}}.
Similarly, the estimated posterior mean of the rating effect ($\gamma_2$) is 
0.0038 and 0.0053 for the 1700-1900 and GM cohorts respectively. For a player with a 
win probability of 0.5 playing as black against an equally rated opponent, playing 
against an opponent rated 100 points lower increases the win probability on average to 
0.63 for GMs and 0.59 for 1700-1900s. 
The increased impact of these effects in higher rated players
seems reasonable, with elite players better able to capitalize on
these advantages.

 
In the 1700-1900 cohort the posterior mean value for $\mu_{\beta}$ is centered around 0, 
indicating that there is little evidence for a global experiential effect. 
The posterior mean estimate for
the GM cohort is positive, although the 95\% interval contains
0, indicating limited weak evidence for a positive experiential
effects.
We also include 
the estimates for each of these parameters for the two intermediate cohorts in 
Figure~\ref{17_19_gm_global}. 
We see a clear linear pattern in the estimates of $\gamma_1$ 
and $\gamma_2$, corresponding to increasing ability in the cohort. However, for all 
cohorts, the estimate of $\mu_{\beta}$ is centered
close to 0.


\begin{figure}[ht]
	\centering
   \includegraphics[width=0.7\textwidth]{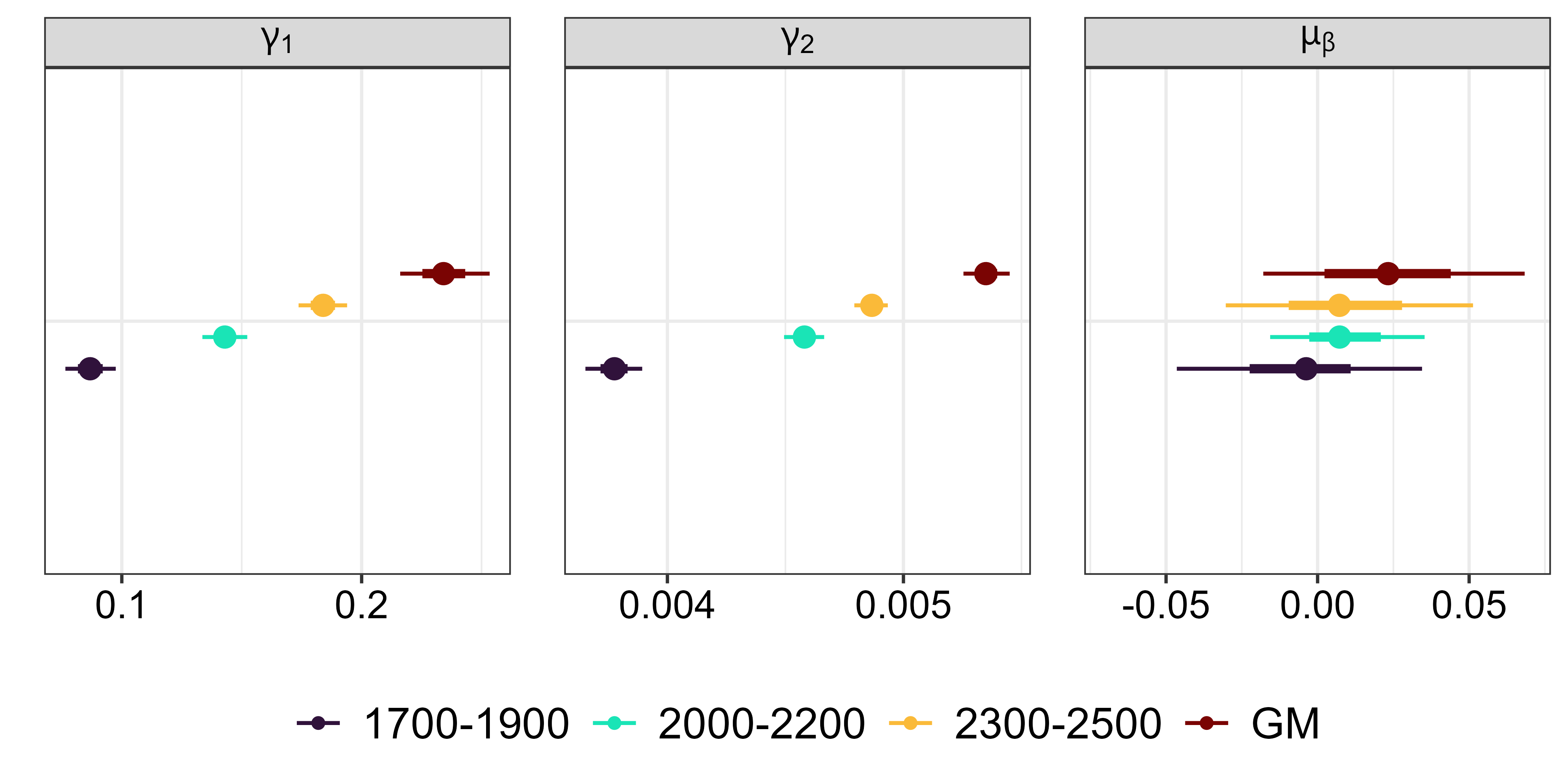}
	\caption{The 66\% and 95\% credible intervals of the estimated colour ($\gamma_1$), rating ($\gamma_2$) and global experiential ($\mu_\beta$) effects in the bullet games.
    The posterior means of the colour and rating effects increase with skill. There is little evidence for a global winner effect in any cohort. } 
	\label{17_19_gm_global}
\end{figure}

While we see little evidence for a global winner-/loser-effect, we also
wish to examine the individual experiential effects and how these vary across
focal players. In Figure~\ref{17_19_gm_winner} 
we take the posterior estimates of $\beta_j$ for each player and
transform them into the change in predicted win probability. 
In particular, if all other parameters were fixed, we compute how much the 
win probability would increase in going from losing the previous game to winning it.


We show 
credible intervals of the posterior estimates of the
change in win probability for 36 focal
players in the 1700-1900 cohort\footnote{Four players in the 1700-1900 cohort were excluded due to not playing any 60+0 bullet games.} and 25 in the GM cohort.
We see that many of these players have changes
close to 0.
The posterior means for this quantity across these two cohorts
range from 
-0.02 to 0.03.
The majority of
95\% posterior intervals contain 0 (or are close to 0).
These effects are small, and indicate small changes in the win 
probability based on performance in the previous game. 
One interesting observation here is that the player with the second smallest estimated
change in the GM 
cohort, the username \texttt{DrNykerstein}, is one of the accounts 
of former world champion Magnus Carlsen. 
While this result seems to
indicate that Carlsen is less likely to win 
the next game if he had won the previous game than other players,
there are several potential factors which could explain this.
One possible explanation concerns data quality. Carlsen 
is widely known for playing in an ``erratic" manner in casual games online, 
and as seen in live streams, may not be taking such games seriously.
In terms of the model used, it is possible that Carlsen 
is different to other players and (for example) has a winner and 
loser effect of different sizes. This would
not be captured by our current model.




\begin{figure}[ht]
	\centering
   \includegraphics[width=0.45\textwidth]{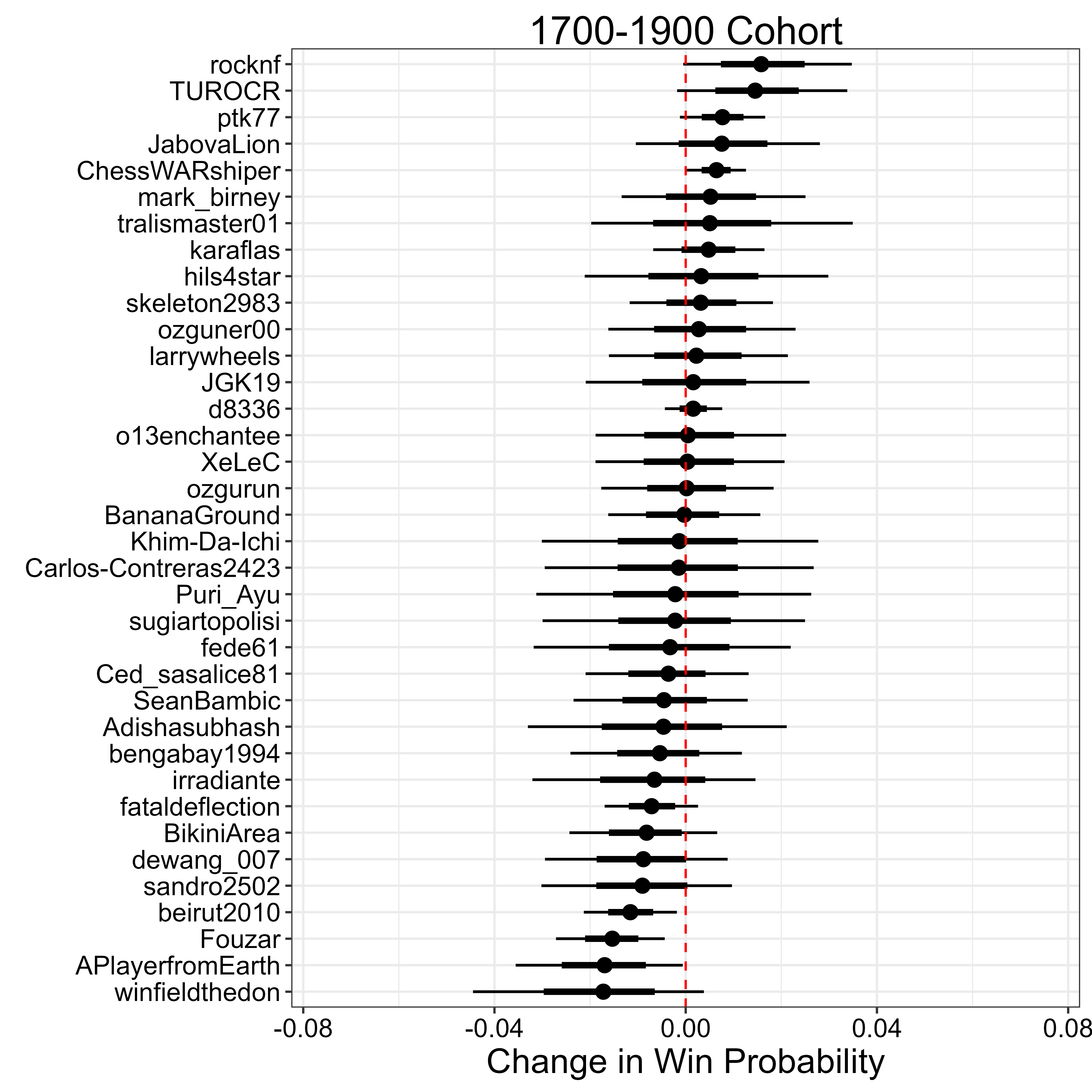}
    \includegraphics[width=0.45\textwidth]{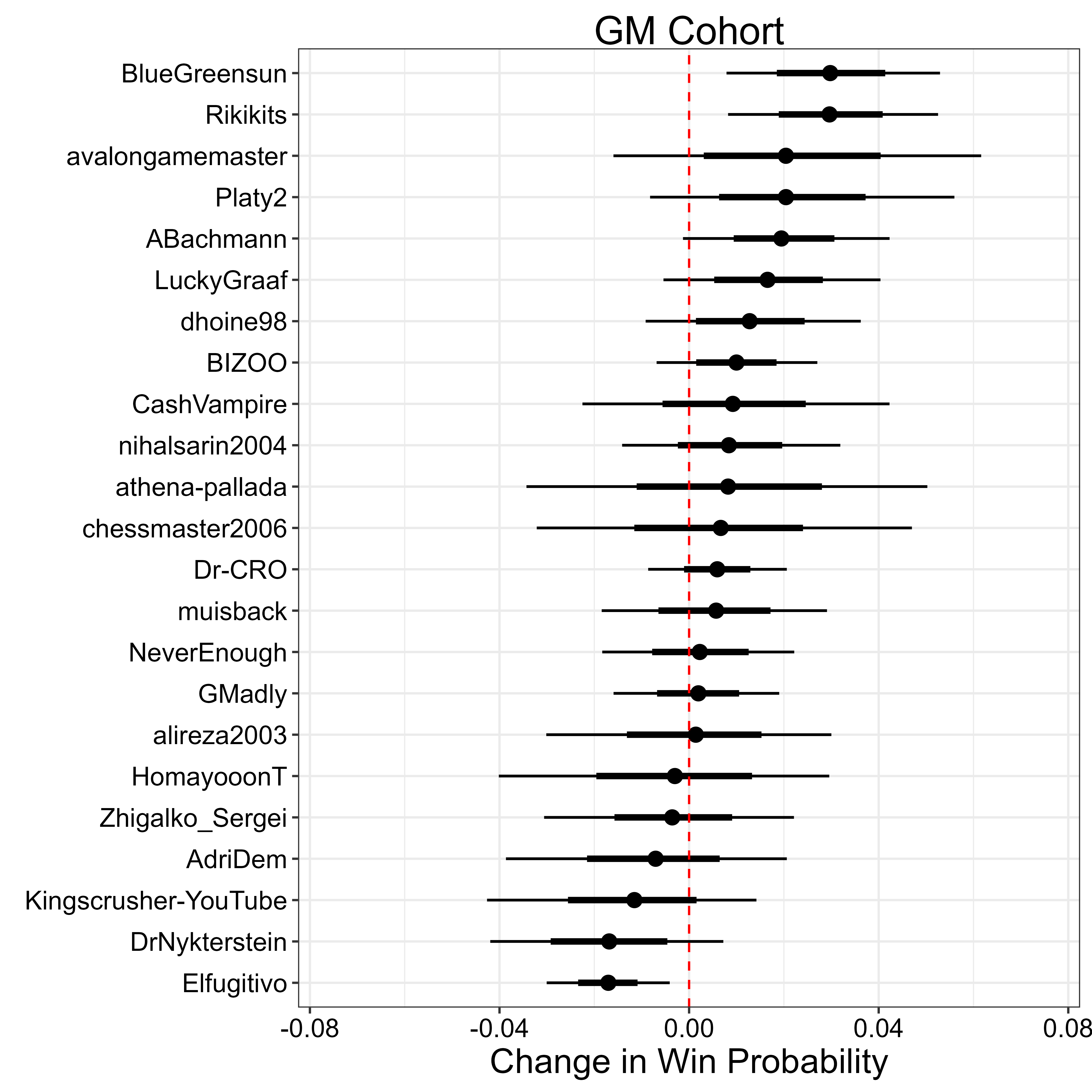}
	\caption{Interpretation of the estimated player-level experiential effects ($\beta_j$ for player $j$)
    for the 1700-1900 Cohort (left) and GM cohort (right). The change in win probability represents the additive boost in win probability a player receives from the experiential effect when changing their previous game from a loss to a win. The 66\% and 95\% credible intervals are shown. Most players exhibit little evidence for non-zero experiential effects, 
    which results in changes in the win probability that are close to zero. 
    }
	\label{17_19_gm_winner}
\end{figure}

\paragraph*{The Use of Different History}

Our proposed model shows little evidence for winner-loser effects above
when $n=1$, which corresponds to the result of the previous game 
alone impacting subsequent performance. However, it is possible
that experiential effects could require longer time periods to develop.
To investigate this we refit the above model now using $n=10$
for computing the win ratio $\tilde{x}_{ij}^{n}$.
It is relatively common for sessions involving more than 10 games 
to be played consecutively, with more than 20\% of all sessions
across both cohorts containing at least 11 games.
In Figure~\ref{gm_comp_global} we show the
posterior estimates for the three key
model parameters for both $n=1$ and $n=10$ for the
GM cohort. We note that 
the posterior distributions for all parameters except
for $\tau_2$ are essentially identical with slightly more uncertainty with $n=10$, due to the
decreased number of sessions.
The posterior mean estimate for $\mu_{\beta}$ 
is closer to 0 when $n=10$, although the intervals show
substantial overlap.
The posterior distribution of $\tau_2$ is 
shifted towards larger values due to the increased variability
of $\tilde{x}_{ij}^{n}$ when $n=10$. Individual winner-loser effects 
(shown in the supplementary material)
show similar patterns with increased variability when $n=10$,
with a similar proportion 
of posterior intervals containing 0 in both cases.
It does not appear that a longer term history provides 
additional evidence for the presence of strong
or persistent experiential effects.

\paragraph*{Comparing Blitz and Bullet Games}
The previous analysis has investigated the presence of experiential effects
in short bullet games lasting at most two minutes. 
To consider the possibility that longer competition is required for such
effects to form, we repeat the previous analysis now using blitz games.
These are games of up to 6 minutes in length. Although these games
are longer, it is still common for successive sequences of games.
We define a focal player starting another blitz game within 7 minutes
of their previous game as games played in the same 
session. With this definition, two thirds of sessions 
played in the GM cohort contain more than one game, 
along with almost three quarters of games in the 1700-1900 cohort.
As described in Section~\ref{sec:data}, we have 
substantially more bullet than blitz games for all players.

We compare the fit of our proposed model with $n=1$
for the GM cohort. The posterior estimates of key parameters 
are shown in Figure~\ref{gm_comp_global_bullet_blitz}.
The estimated fixed effects $\gamma_1,\gamma_2$
show larger variability in the smaller blitz dataset.
In both cases, we see the distribution of the estimated global
winner-/loser-effect is close to zero, with significantly more variability
in the smaller Blitz dataset.




\begin{figure}[ht]
	\centering
    \begin{subfigure}{0.45\textwidth}
        \centering
        \includegraphics[width=\textwidth]{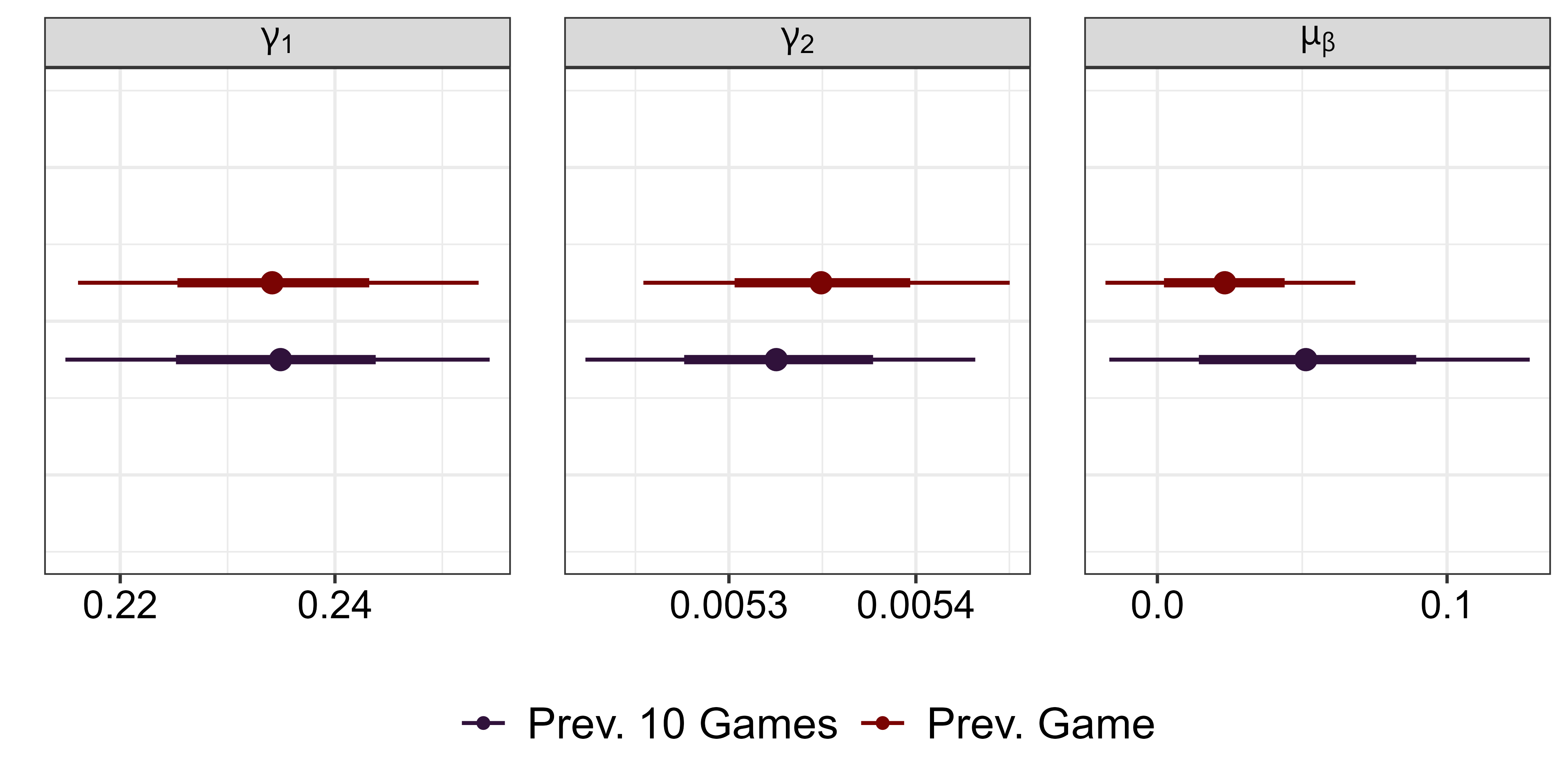}
        \caption{Comparing the estimated parameters when $n=1$ vs $n=10$.}
        \label{gm_comp_global}
    \end{subfigure}
    \begin{subfigure}{0.45\textwidth}
        \centering
        \includegraphics[width=\textwidth]{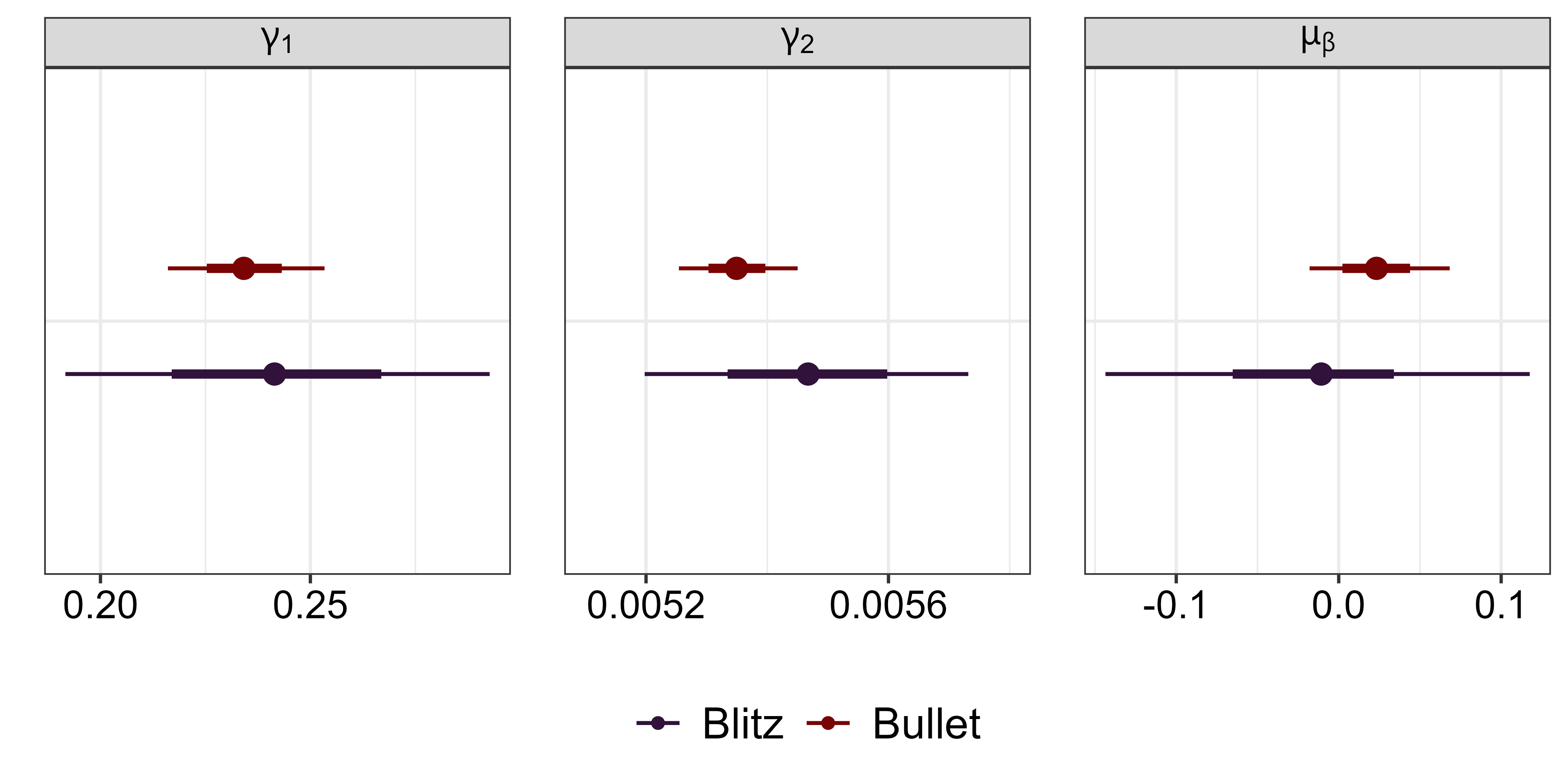}
        \caption{Comparing the estimated parameters using bullet or blitz data.}
        \label{gm_comp_global_bullet_blitz}
    \end{subfigure}
	\caption{Comparison of the estimated fixed and global effects in varying settings for the GM cohort.
    In each case we show 66\% and 95\% posterior intervals.}
	\label{gm_comp}
\end{figure}

In Figure~\ref{gm_comp_winner_bullet_blitz} we show the estimated
individual winner-loser effects for the members of the GM cohort
using both the blitz and bullet data. Some of the distributions
appear to shift slightly with the blitz data but all
demonstrate larger variability using the blitz data.
This is expected, as all play more bullet games (on average 9000 bullet games compared to 
1300 blitz games in this data, with an ever bigger difference in other cohorts).

\begin{figure}[ht]
	\centering
   \includegraphics[width=0.65\textwidth]{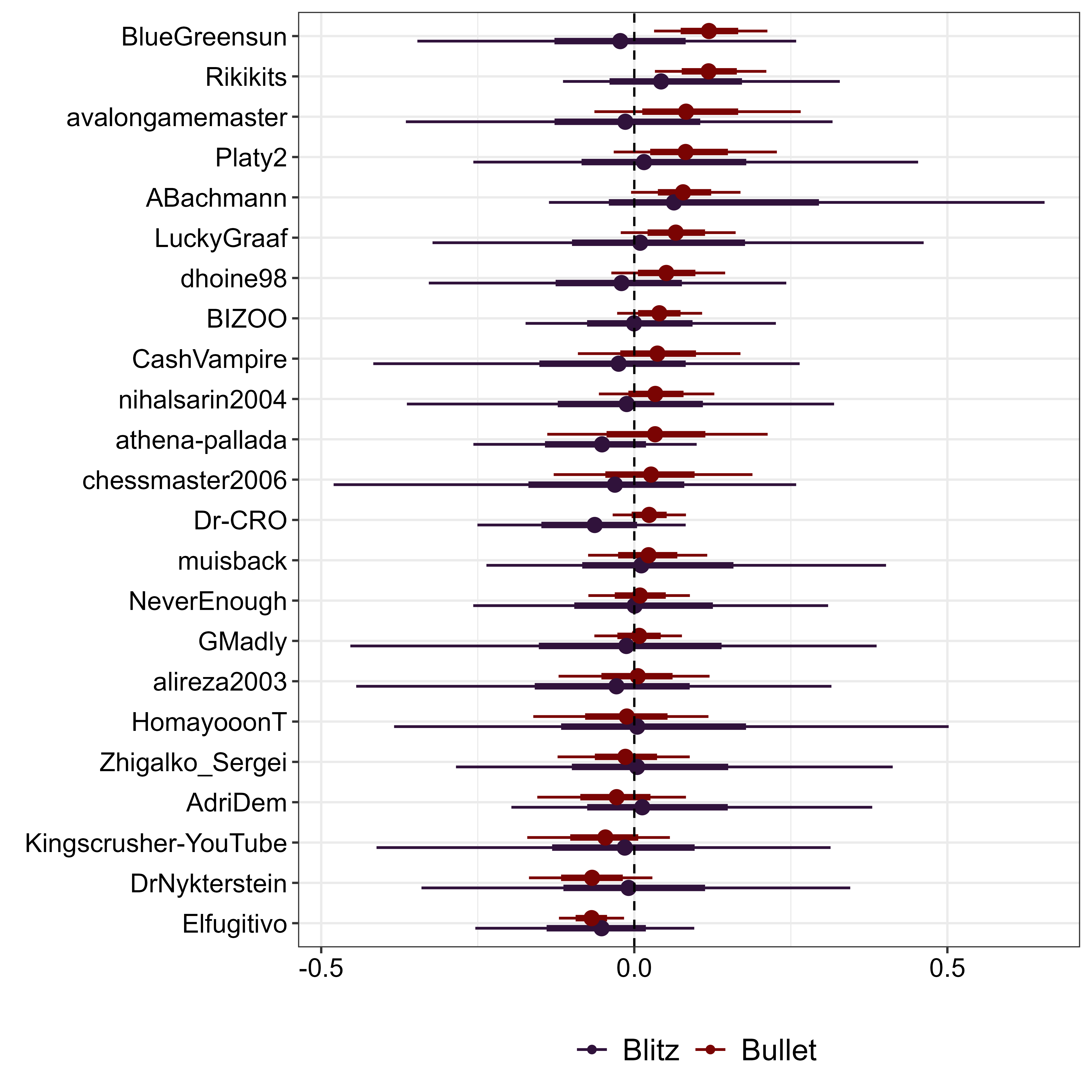}
	\caption{Winner-loser effects for the GM data,
    from fitting the proposed model to Bullet and Blitz data. 
    We see considerably larger variability with the smaller blitz dataset. }
	\label{gm_comp_winner_bullet_blitz}
\end{figure}

Overall, we see little evidence for consistent experiential
effects from fitting the proposed Bayesian logistic regression model to
large volumes of online chess data. Across a variety of formats and
time scales for the formation of these effects, we do not see any
evidence for such effects across the population of players 
considered. While there is some evidence for individual players
displaying winner-loser effects, 
and some evidence of a small global effect 
in the highest rated players, these effects appear to be weak.
An important concern with the current analysis, investigating
these effects in the context of recent performance, is the 
potential for high temporal correlation between successive games.
In Section~\ref{sec:checking} we consider this and provide several
tools to validate our model.

\section{Model Validation}
\label{sec:checking}

The previous sections demonstrated that, under the proposed
hierarchical regression model, there is little evidence 
for a consistent winner-/loser-effect across the population
of online chess players studied. However, our model 
is very much a simplification of the complex nature of such data.
One important consideration which our model does not account 
for is the temporal nature of this data. 
Many of the focal players in these datasets have several
years of games in the dataset. It is possible that 
potential experiential effects can change over time, 
which we would fail to capture in this model.
There is the potential for strong temporal
dependence in this data which we do not account for in our 
simplified model. 
Below we consider one method to evaluate the fit 
of our proposed model in the context of these potential issues. We
include an additional model checking procedure in the appendix, 
{\color{red}examining permutation distributions
of the game results.}
We also include there a simulation study, indicating 
roughly how many games would be required to detect experiential effects of varying sizes.
We demonstrate that while our model may somewhat suffer 
from ignoring the clear temporal structure in 
this data, the current model can capture much of the
dependence in this data.


\subsection*{Posterior Predictive Checks}

Given the Bayesian model used here, a natural tool to 
assess our proposed model is to utilise posterior predictive checking. 
This is a widely used procedure where replicated data under the fitted Bayesian
model is compared to the original observed data \citep{gelman2013bayesian}.
Following the notation of \cite{gelman2013bayesian} we define 
the posterior 
predictive distribution as
\begin{equation}
    p(y^{\text{rep}}|y) = \int p(y^{\text{rep}}|\theta)p(\theta|y)d\theta, 
\end{equation}
where $\theta$ is the vector of parameters in the model.
We can then draw replicated data from this distribution, namely we can simulate the 
result of a given match, given the covariates 
and each posterior draw of the model parameters.
A failure of these replications to match the observed 
data can identify potential flaws in the model. These posterior predictive checks
are often done using some generated quantity from the original dataset.
Here, a natural quantity to consider is the number of games won by a
focal player in the observed data. An extension of this is to consider
the generated quantity corresponding to the rating evolution of a player 
over replicated datasets, simulating each game
result and then using these outcomes to construct the game history for future
games.
This is the quantity we will use for posterior predictive 
checking below.

Given the size of our original dataset, creating thousands of 
posterior predictive replications of each game has large memory requirements.
Instead, we will restrict our data to a selection of focal players in 
the GM cohort.
For these players we utilise their 2000 most recent bullet games in the dataset.
We will use the first half of this data to estimate the posterior 
distribution, before then simulating the results of the games in the second half, 
using each of the draws from the posterior and the
value of $\tilde{x}_{ij}^{n}$ based on the simulated 
result of the previous game with this set of
posterior draws. For each set of parameter draws from 
the posterior we can then compute the estimated ranking under 
a Glicko-2 rating system, similar to that used by Lichess.
We use default values for the constant parameters in this 
model, which gives similar results to the (unknown) parameters used by Lichess.
This gives us posterior predictive draws of the rating of 
a player over 1000 games, which we can then compare to the true evolution
of these scores.

We fit our previous proposed model with $n=1$ on the previous 1000 games before then constructing a posterior predictive distribution
of the Glicko-2 rating of each player using
the final 1000 games 
for the GM cohort. As we 
use the standard number of chains (4) and sampling iterarions (1000) in CmdStanR,
this gives us 4000 posterior draws for each parameter.
We then use each set of draws to simulate the results of each game in the last 
1000 games. This gives us 4000 replications of the 1000 games played by each 
focal player and the corresponding evolution of their Glicko-2 score.
We show these replications for 2 GMs in Figure~\ref{fig:gm_ppc}.
While these {\color{red}replications} show large uncertainty we see that they capture the overall trend
well, in particular capturing the steady increase in rating shown in Figure~\ref{ppc_elo_nihal}.
This indicates that our model is able to predict the future temporal evolution of
the ability of players well.
We include an additional model checking 
{\color{red}approach, examining the results of our model when we permute the game results}
in the appendix, along with this posterior predictive check 
for players from each of the other cohorts.

\begin{figure}[ht]
	\centering
    \begin{subfigure}{0.45\textwidth}
        \centering
        \includegraphics[width=\textwidth]{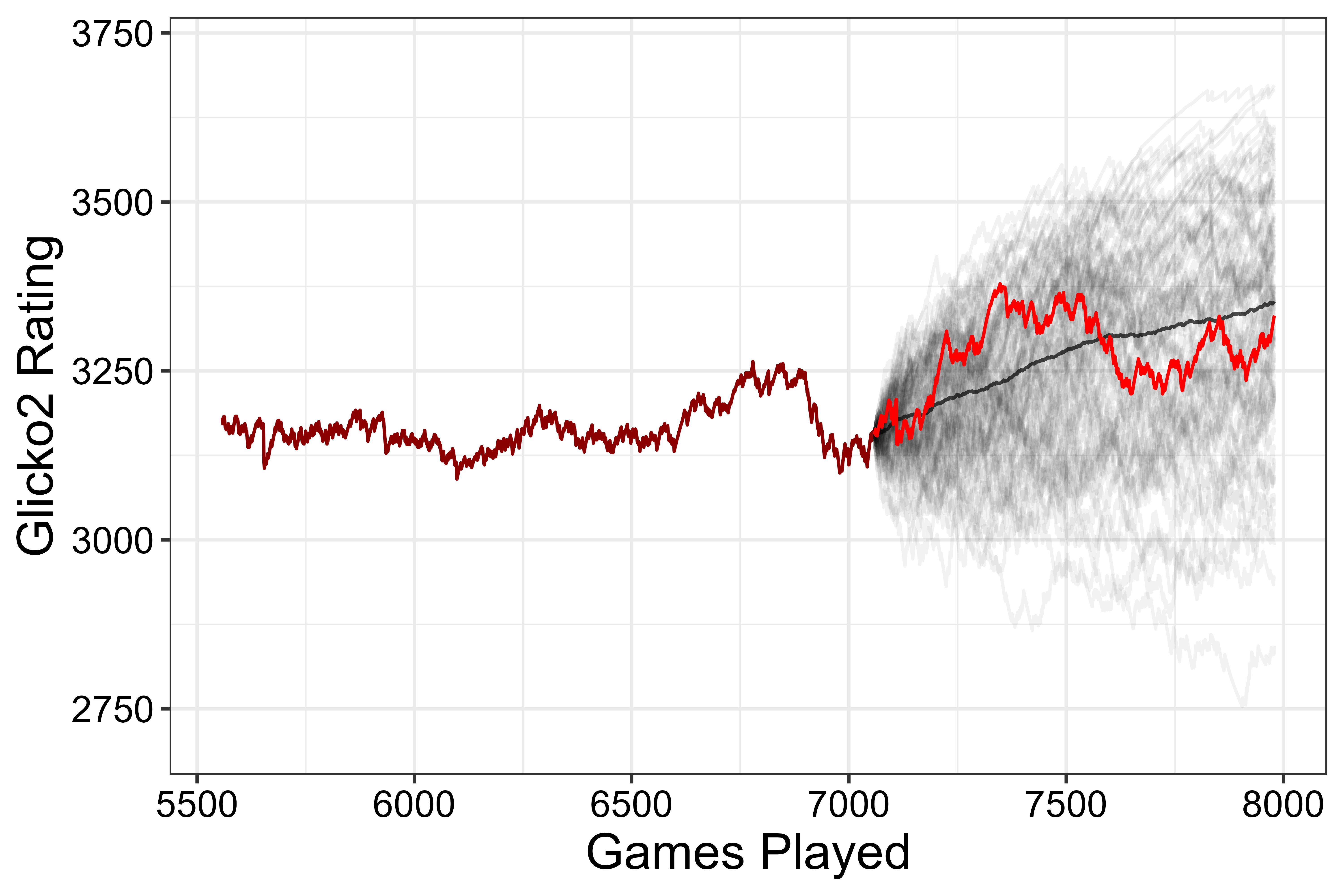}
        \caption{Posterior predictive distribution for Magnus Carlsen.}
        \label{ppc_elo_magnus}
    \end{subfigure}
    \begin{subfigure}{0.45\textwidth}
        \centering
        \includegraphics[width=\textwidth]{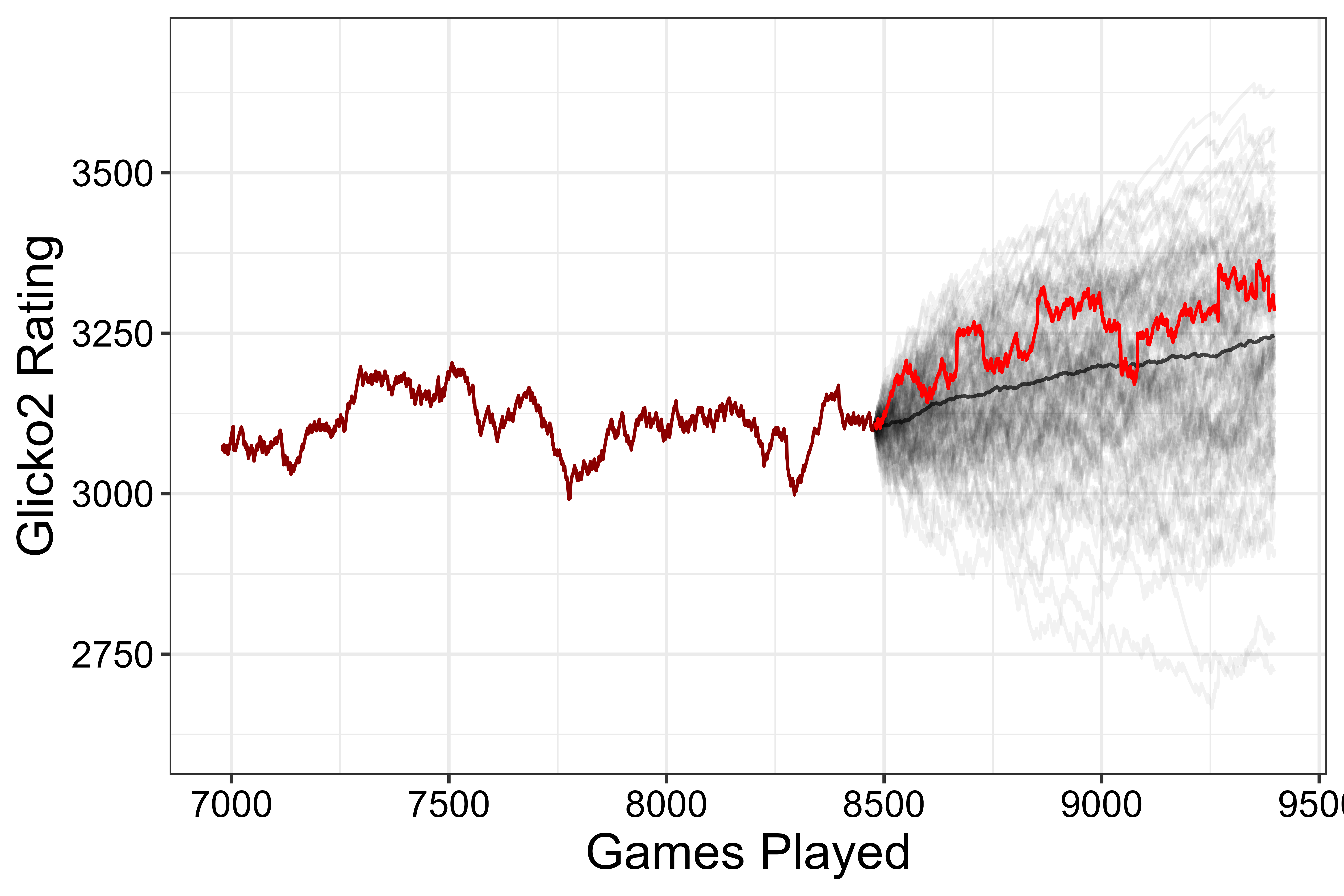}
        \caption{Posterior predictive distribution for Nihal Sarin.}
        \label{ppc_elo_nihal}
    \end{subfigure}
	\caption{Posterior predictive distribution of Glicko-2 ratings for 2 GMs. We show the true Lichess rating in red, with 4000 draws of the predicted rating evolution. The mean of these draws is shown in the solid black line.}
	\label{fig:gm_ppc}
\end{figure}


\section{Summary, Future Directions}
\label{sec:summary}

In this work we have proposed a Bayesian hierarchical logistic regression 
model to investigate the presence of experiential effects in online chess.
Having fit this model to cohorts of players of varying ability
we find that there is no strong evidence for the presence
of global winner/loser-effects, as commonly defined
in the literature. However, a small proportion of individual players from all cohorts show evidence of possessing such effects. This is 
consistent across multiple formats of chess and 
the number of previous games that this effect could persist over.
Our model is a general Bayesian model which can easily
be modified to other data of this form, providing a tool for estimating
both individual and population level experiential effects.
While we do not believe that our model perfectly describes the dynamics 
of chess, our model checking procedures indicate that our model
appears reasonable and captures many of the key characteristics of
this data. Our results indicate that a more flexible model
may be required for players of higher ability, 
such as GMs, where we saw some limited evidence for $\mu_{\beta}>0$.
This would need to be investigated further, and could
be related to the larger impact of the covariates (color and rating difference)
in games between better players. Another concern which arises when examining these top players is that the opponent pool is significantly smaller. As such, the assumptions
of playing random opponents may not {\color{red} be} appropriate for particularly strong players.
Modifying our approach to account for non-random opponents is an important 
future extension, {\color{red} along with examining the reliability
of the history covariate we use here.}
{\color{red}A key advancement of the Glicko-2 rating system is the ability to 
incorporate individual rating deviation when examining the rating difference between
players \citet{glickman2012example}.
This gives a more reliable measure of the difference in ability 
between opponents, however this rating deviation is not available
from the Lichess API. Being able to obtain and use this deviation 
could significantly improve our model.}

There are many other potential future directions which could expand on this work. 
In shorter chess games it is common to attempt to ``flag" your opponent,
aiming to beat them by forcing them to run out 
of time. A large proportion of bullet games are won this way 
(24\% of bullet games in the GM cohort)
and results of this form may have a different impact on 
future performance than other
endings. We have also not attempted to separate out 
winner and loser effects or allow them to have different
magnitudes in the current formulation.
The current model also does not consider draws (which are rare in
this data) and 
the potential impact they could have on future performance.
Similarly, while we consider Bullet and Blitz games separately here 
we could also use both for predicting the outcome of either format,
although it is very uncommon for players to switch between formats 
in a given session.
{\color{red}More broadly, given the general nature of the model
proposed here, we hope to also apply it to recent large animal interaction datasets,
where the presence of experiential effects remains hotly debated.}


\begin{funding}
  Adam Gee acknowledges the support of an NSERC CGRS M scholarship
  and Owen Ward acknowledges
  support from an NSERC Discovery Grant RGPIN-2023-05335.
\end{funding}

\begin{appendix}
\renewcommand{\thefigure}{A\arabic{figure}}
\setcounter{figure}{0}

\section*{Appendix: Additional Analyses}
Here we include some additional model fitting results which were omitted from the main text.
In particular, we show selected model estimates and model diagnostics
for the other cohorts considered along with an additional
model checking procedure and simulation study.
We also investigate the effect of using varying numbers of games for the historic influence.

\paragraph*{Examining Winner-/Loser-Effects across Cohorts}
In Figure~\ref{fig_s1} we show
the posterior distributions of individual $\beta$ parameters
for each player in the 2000-2200 and 2300-2500 cohorts. 
In each case we see similar results to those shown in the main text. The majority of posterior 
distributions for $\beta_j$ contain zero, with a small number of players showing positive or negative 
estimates. This behaviour is common across all ranges of rating we considered.

\begin{figure}[ht]
	\centering
    \begin{subfigure}{0.45\textwidth}
        \centering
        \includegraphics[width=\textwidth]{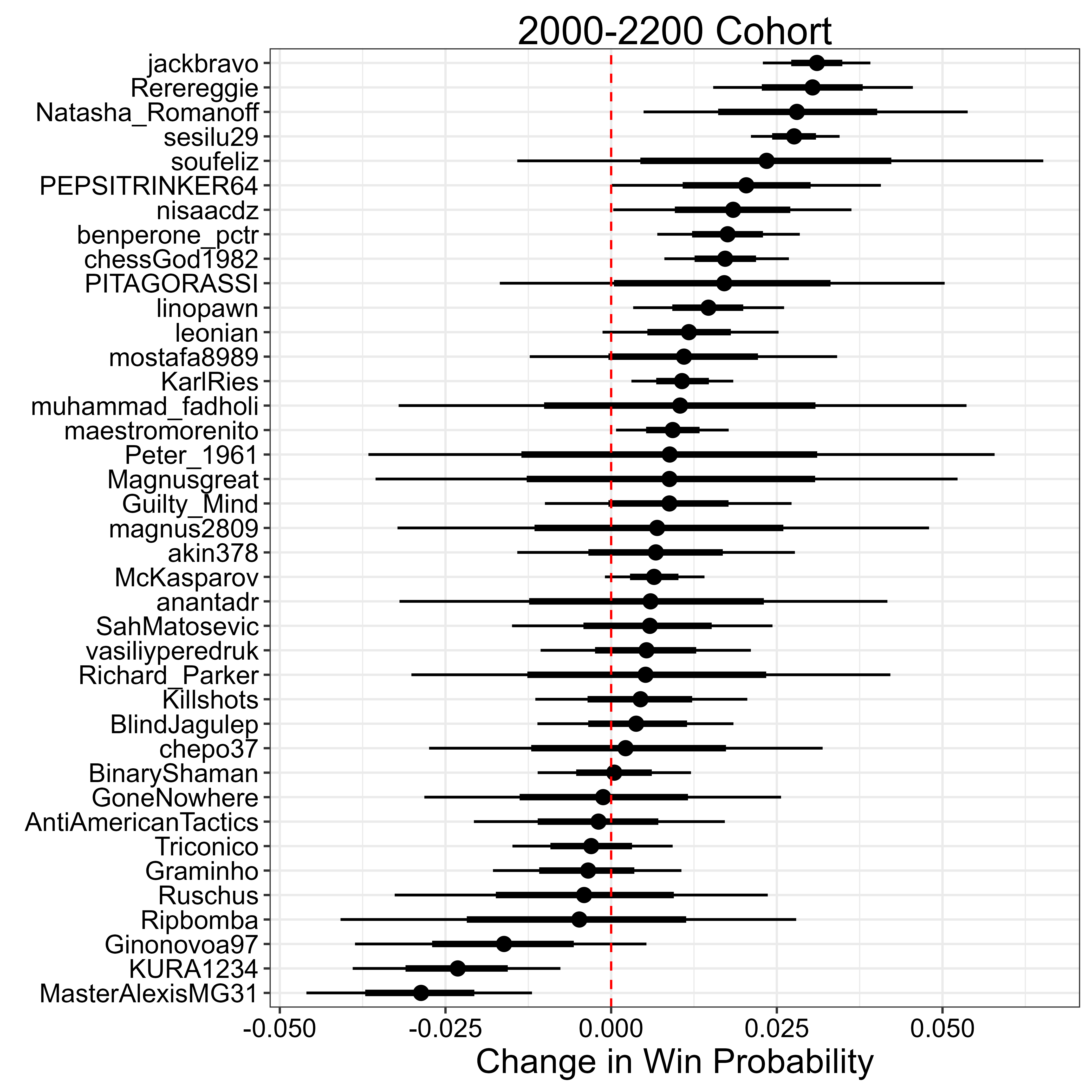}
        \caption{Players in the 2000-2200 cohort.}
        \label{fig_s1_left}
    \end{subfigure}
    \begin{subfigure}{0.45\textwidth}
        \centering
        \includegraphics[width=\textwidth]{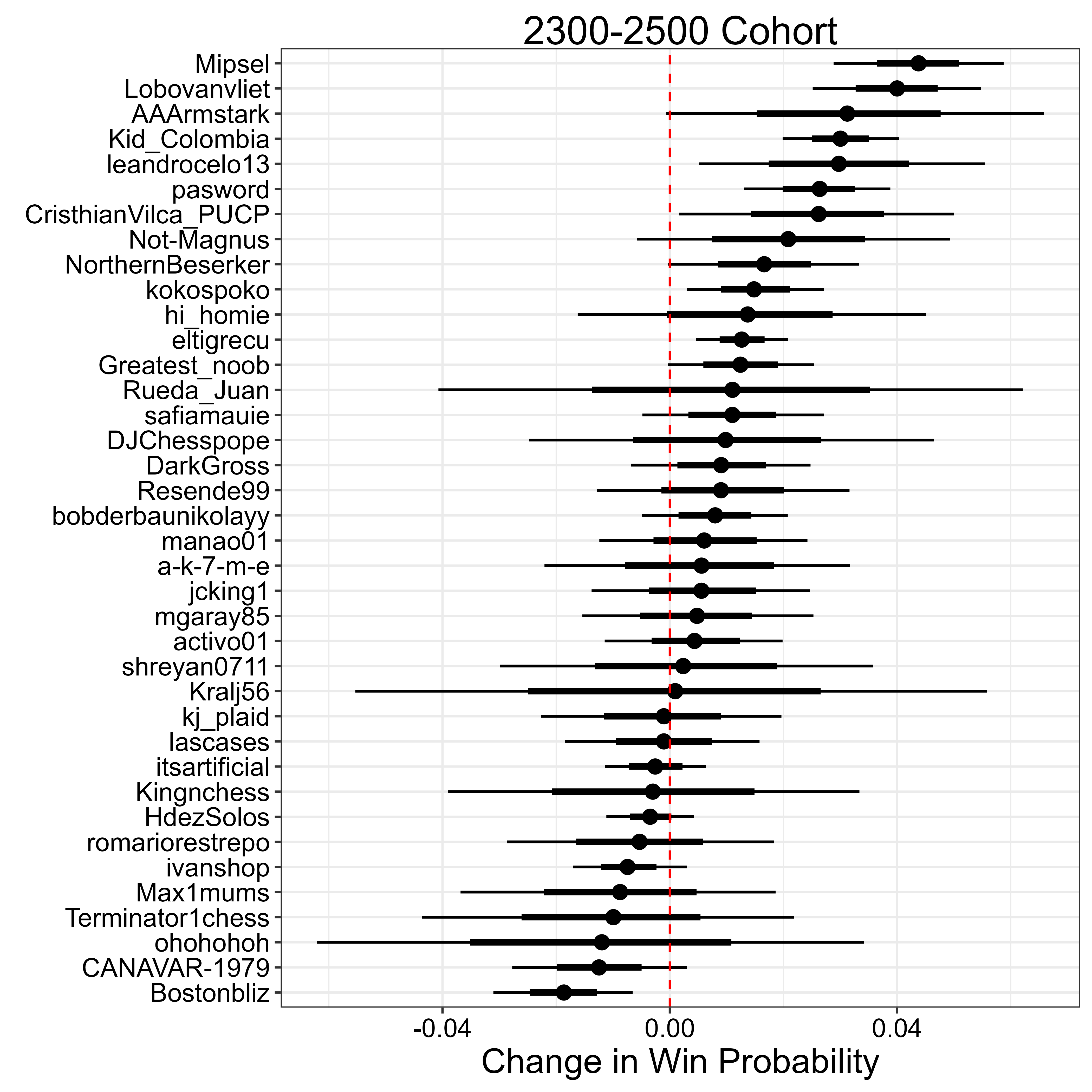}
        \caption{Players in the 2300-2500 cohort.}
        \label{fig_s1_right}
    \end{subfigure}
	\caption{The estimated change in win probability for the other two cohorts we consider.  }
	\label{fig_s1}
\end{figure}

\paragraph*{Posterior Predictive Checking for Other Cohorts}
In Figure~\ref{fig_s2} we demonstrate further posterior predictive checks
using players from cohorts not included in the main text. 
We again construct the posterior predictive distributions
of the Glicko-2 rating evolution. We show sample players in the 1700-1900, 2000-2200 and 2300-2500 cohorts. In each case we see that the predictive distribution 
appears to match the general pattern over future games, indicating
our model is able to predict the temporal nature of
the ranking evolution.

\begin{figure}[ht]
	\centering
    \begin{subfigure}{0.32\textwidth}
        \centering
        \includegraphics[width=\textwidth]{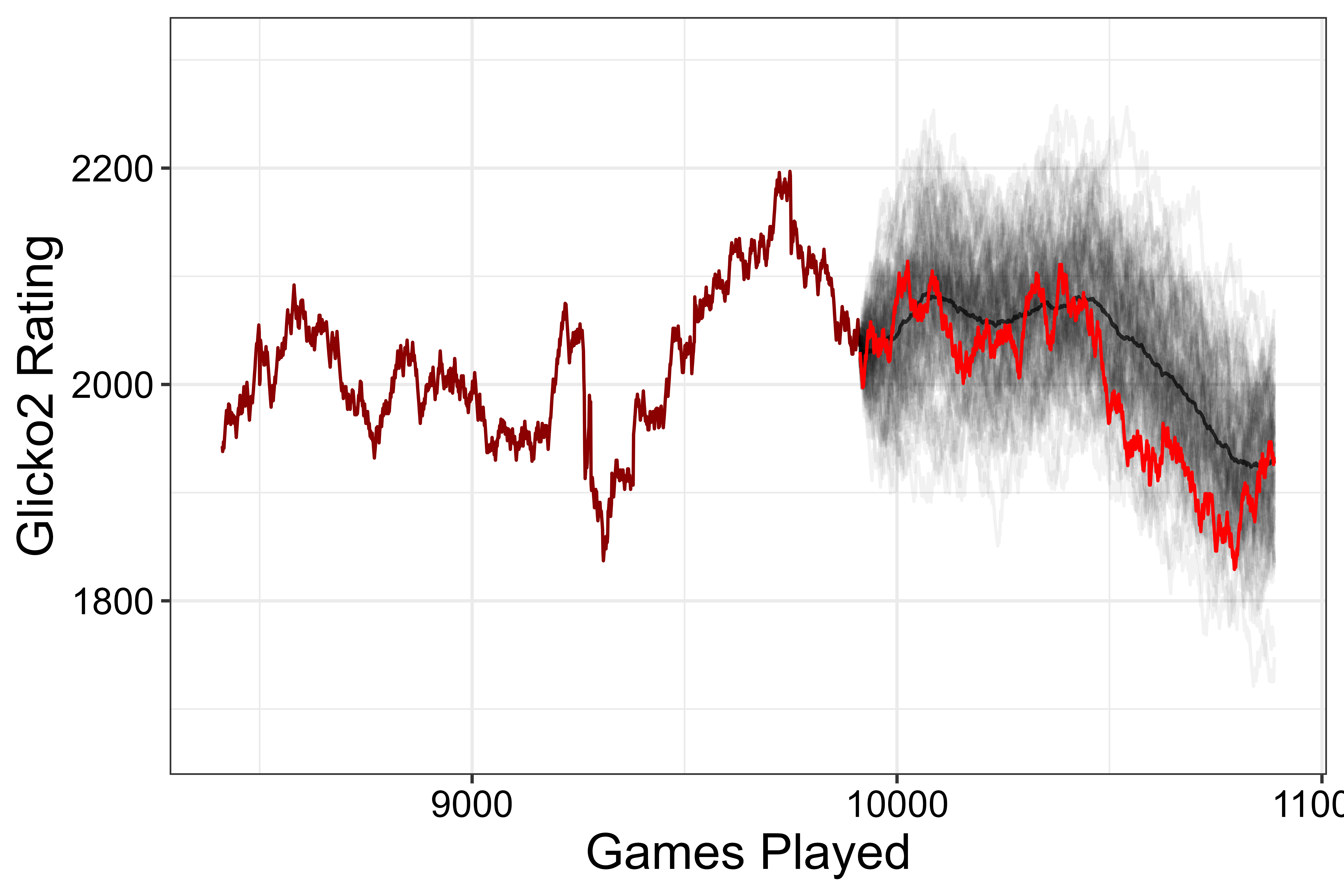}
        \caption{Posterior predictive distribution for "APlayerfromEarth" (1700-1900 Cohort).}
        \label{fig_s2_left}
    \end{subfigure}
    \begin{subfigure}{0.32\textwidth}
        \centering
        \includegraphics[width=\textwidth]{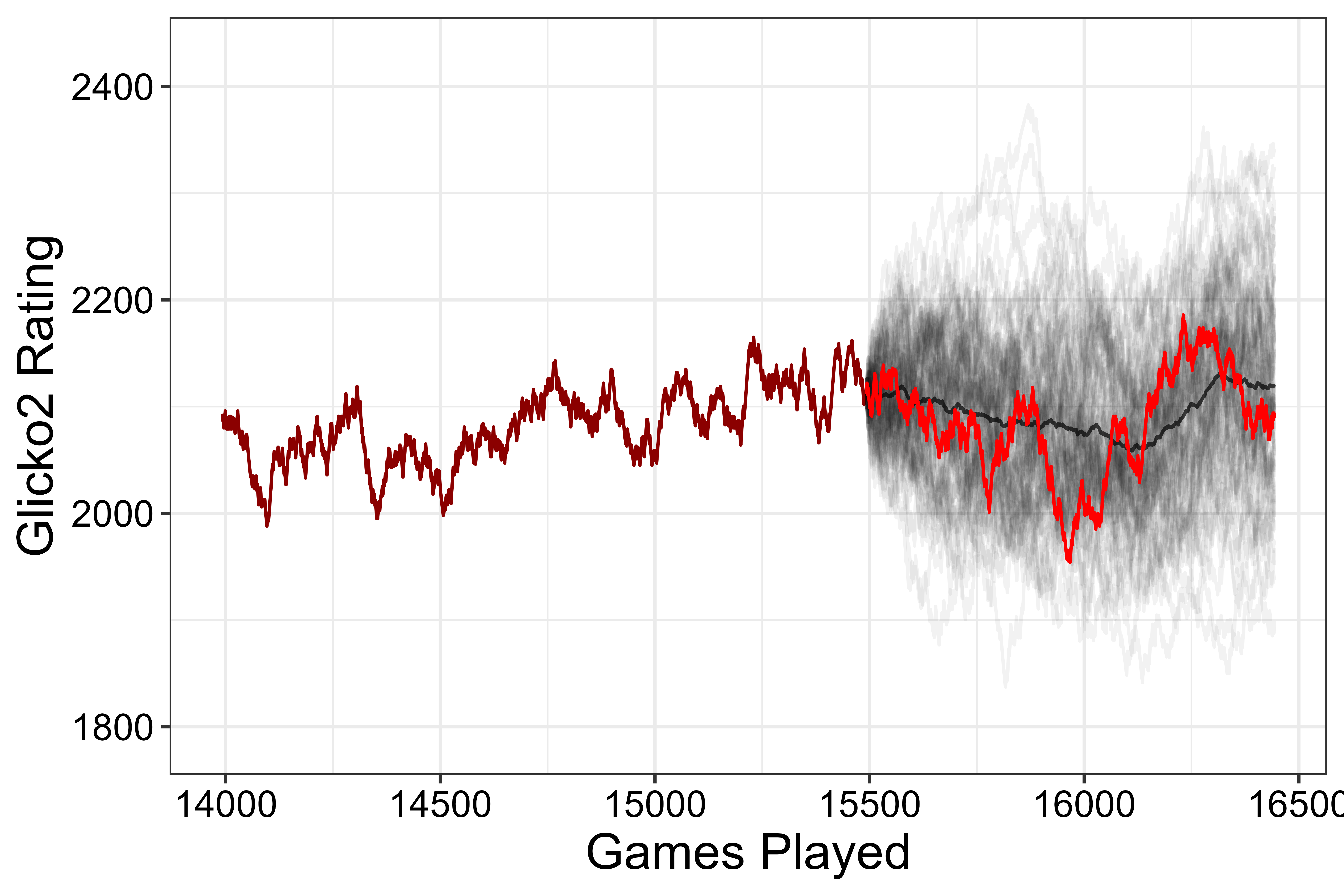}
        \caption{Posterior predictive distribution for "Rerereggie" (2000-2200 Cohort).}
        \label{fig_s2_middle}
    \end{subfigure}
    \begin{subfigure}{0.32\textwidth}
        \centering
        \includegraphics[width=\textwidth]{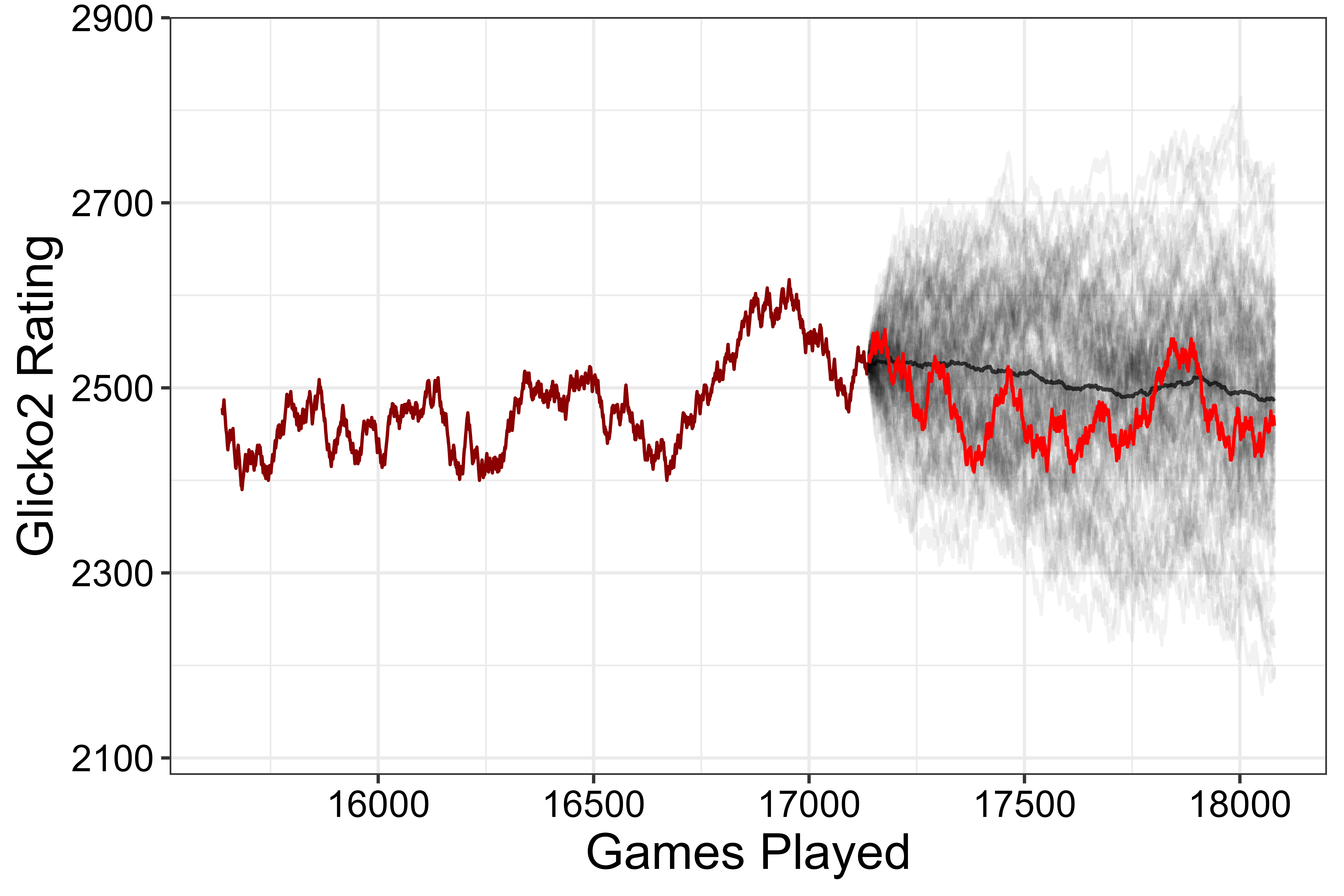}
        \caption{Posterior predictive distribution for "Lobovanvliet" (2300-2500 Cohort).}
        \label{fig_s2_right}
    \end{subfigure}
	\caption{Posterior predictive distribution of Glicko-2 ratings for 3 players in different rating cohorts. We show the true Lichess
 rating in red, with 4000 draws of the predicted rating evolution.
 The mean of these draws is shown in the solid black line. }
	\label{fig_s2}
\end{figure}

\paragraph*{Investigating the Role of $n$}
In the main text we fit our model with $n=1$ and $n=10$, considering the previous 
performance over the previous game and the previous 10 games respectively.
Given that many playing sessions contain less than 10 games, a natural
question is whether $n=10$ is too long a time scale for potential
experiential effects to persist.
In Figure~\ref{fig_s3} 
we show the key global parameters $\gamma_1,\gamma_2,\mu_{\beta}$
for the 1700-1900 and GM cohorts, for $n=1,5,10$.
In each case we see that the estimates of these
parameters are essentially identical, which is also the case for the 
individual parameters $\beta_j$.

\begin{figure}[ht]
	\centering
    \begin{subfigure}{0.45\textwidth}
        \centering
        \includegraphics[width=\textwidth]{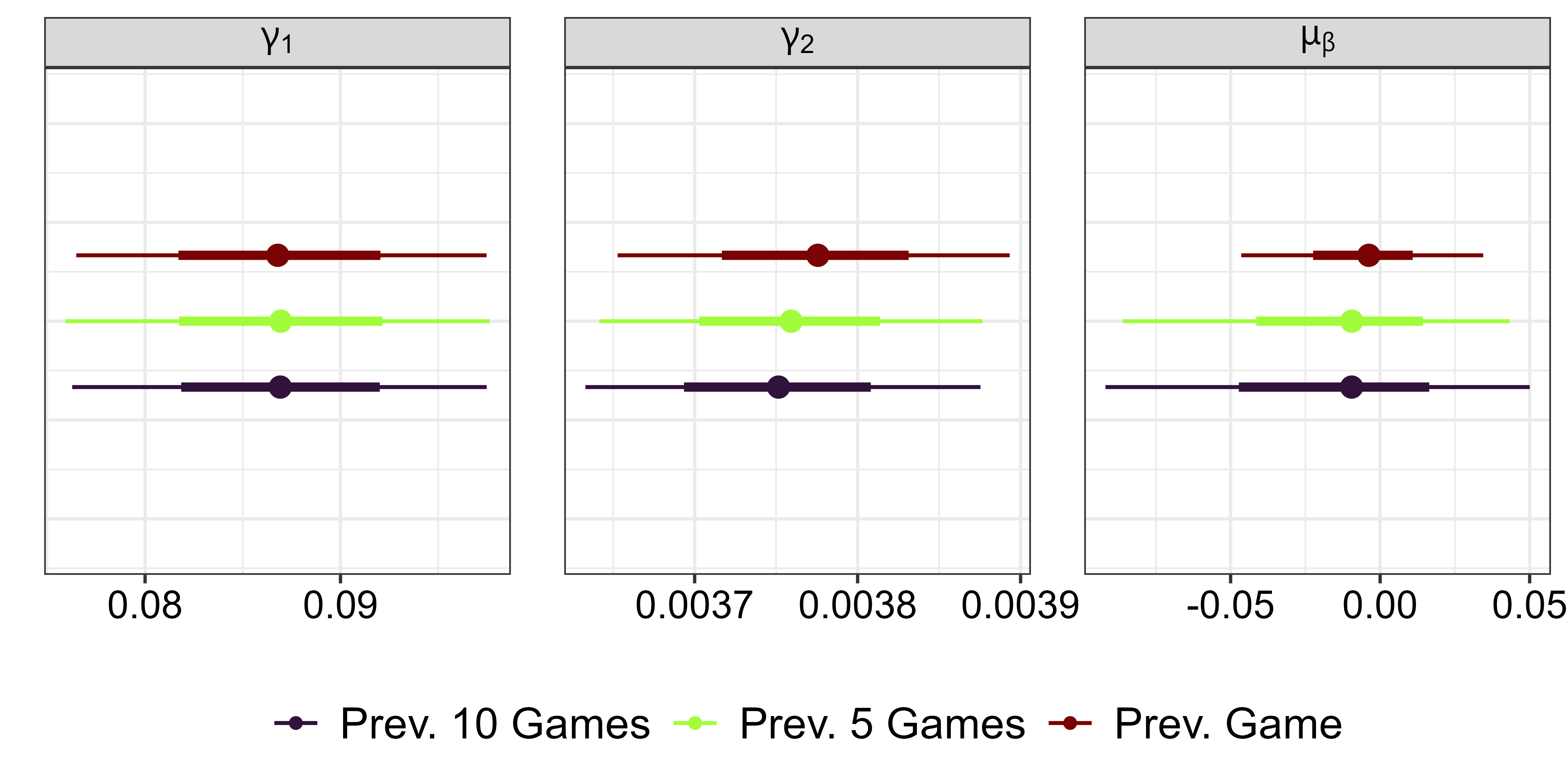}
        \caption{1700-1900 Cohort}
        \label{fig_s3_left}
    \end{subfigure}
    \begin{subfigure}{0.45\textwidth}
        \centering
        \includegraphics[width=\textwidth]{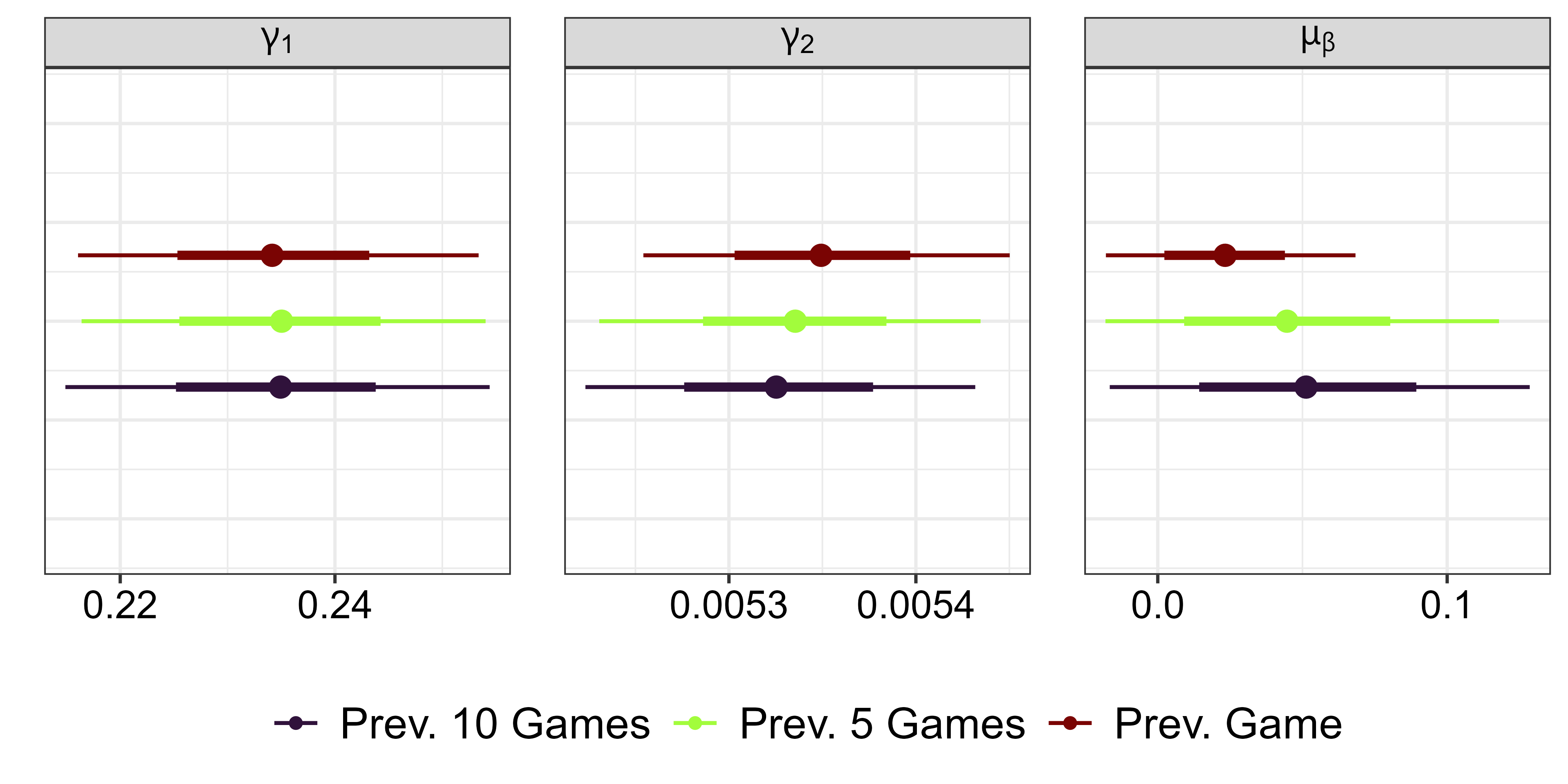}
        \caption{GM Cohort}
        \label{fig_s3_right}
    \end{subfigure}
	\caption{Posterior estimates for the key global parameters $\gamma_1,\gamma_2, \mu_{\beta}$
    as we vary $n$ in our fitted model.}
	\label{fig_s3}
\end{figure}

\paragraph*{Permuting the Game Results}
As an alternative approach to assessing the robustness of our model, we consider permuting the data. A similar approach was considered by \citet{ding_hockey}, to examine evidence for a hot hand effect in hockey. For computational considerations, we perform this analysis with a subset of the complete bullet data. We use 
20 players from each of the 1700-1900 and GM cohorts,
taking the final 1000 games from each. We randomly permute the results within each focal player's data 1000 times, giving us 1000 permuted datasets. For each of these we fit our proposed model with $n=1$. We then examine the distribution of the posterior means of
the experiential effects that are generated from each of these model fits. 

Figure~\ref{fig:gm_perm} shows these ``permutation distributions'' for 20 players in the 1700-1900 cohort and 20 players in the GM cohort, with the central 66\%
and 95\% intervals for each estimate of $\beta_j$ under these permutation
distributions.
{\color{red} 
For all players the true estimated experiential effects lie within these 95\% permutation intervals.
This aligns with our previous results, indicating that individual experiential effects
are small and close to zero.}
We note that when we repeat this analysis for the fixed effects $\gamma_1$ and 
$\gamma_2$ the distribution of these effects under the permutations is clearly different 
from the estimated effects, for all cohorts, as would be expected.

\begin{figure}[ht]
	\centering
    \begin{subfigure}{0.45\textwidth}
        \centering
        \includegraphics[width=\textwidth]{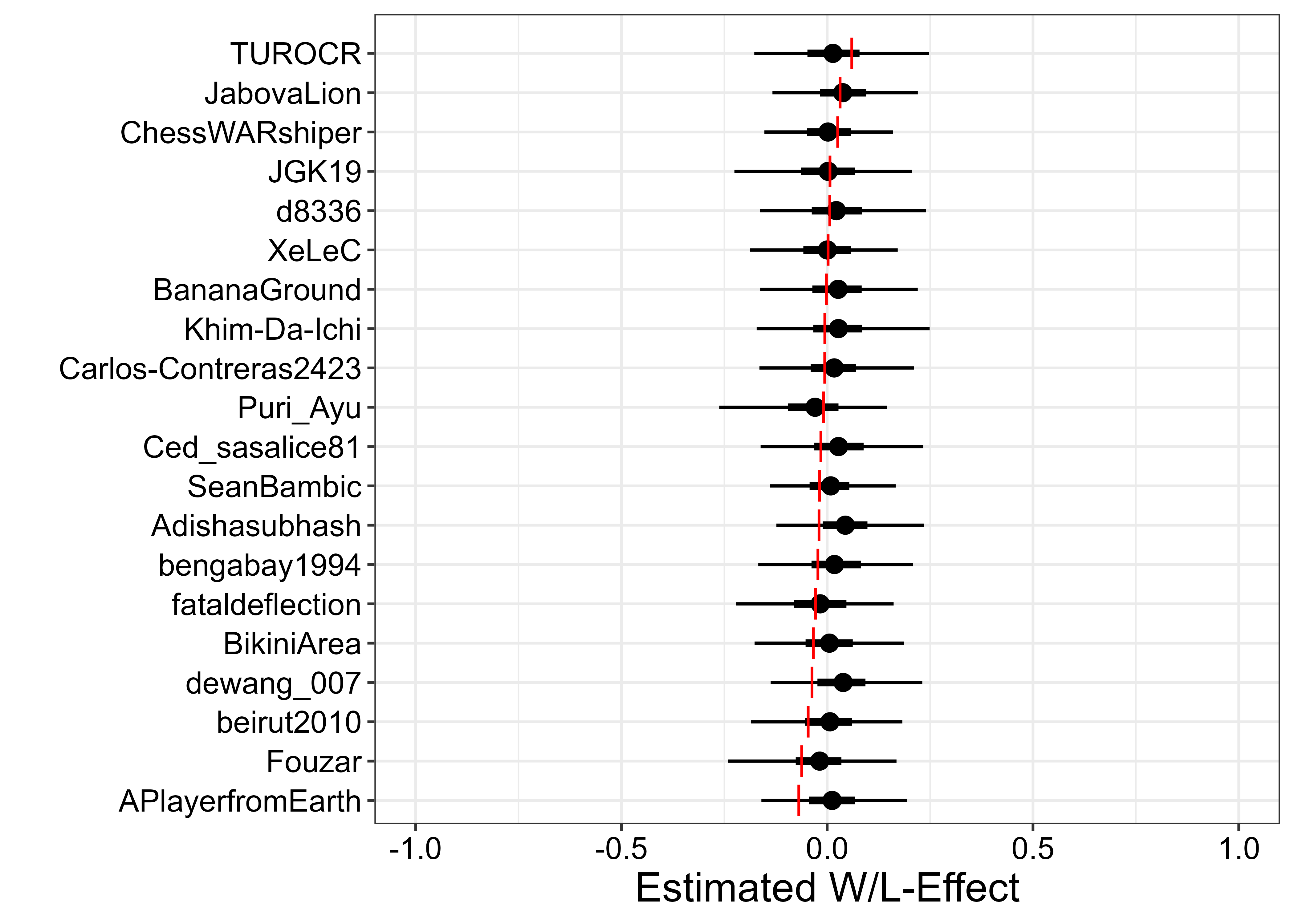}
        \caption{Permutation estimates
    for 20 players in 1700-1900 Cohort.}
        \label{fig:perm_17_19}
    \end{subfigure}
    \begin{subfigure}{0.45\textwidth}
        \centering
        \includegraphics[width=\textwidth]{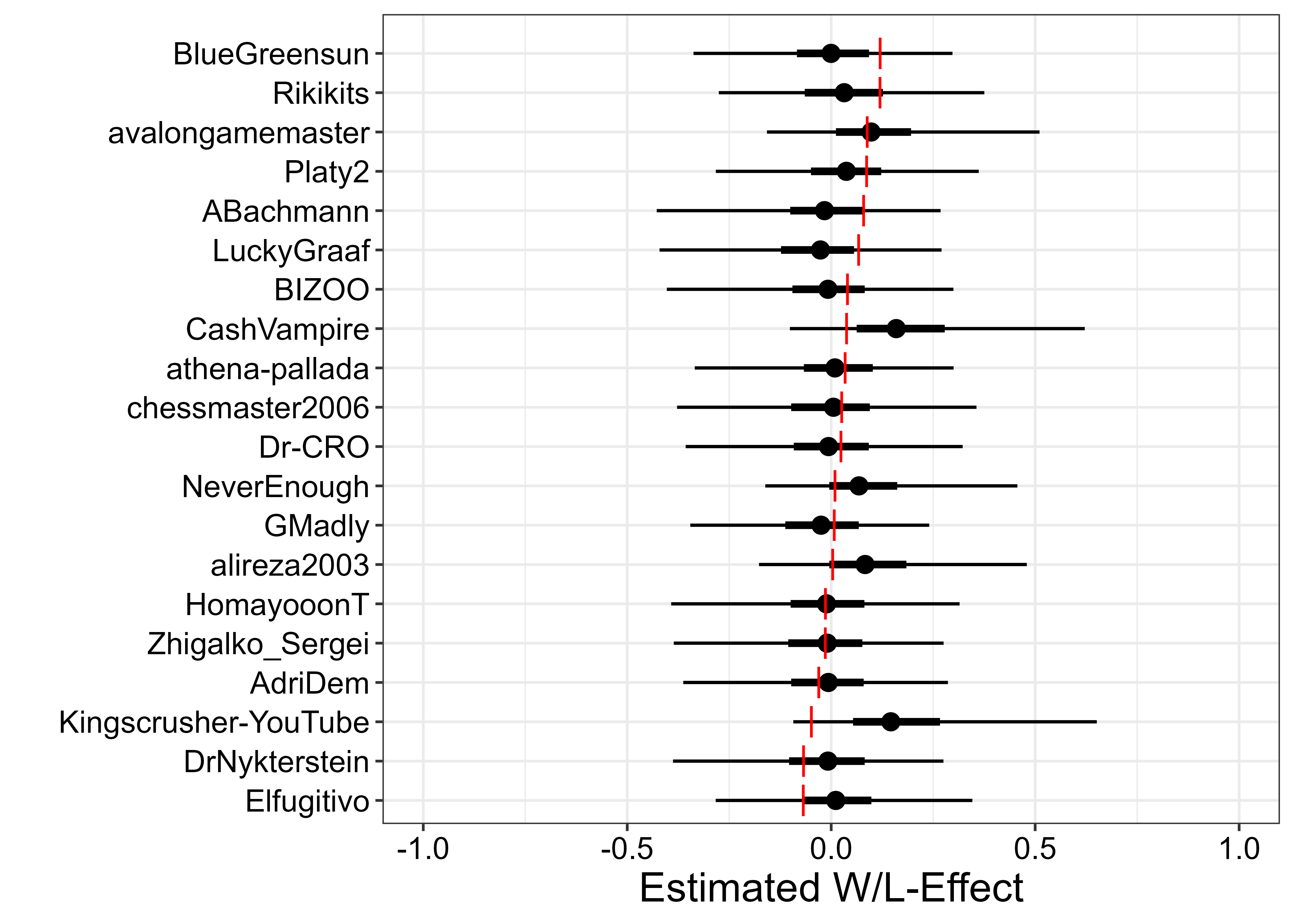}
        \caption{Permutation estimates for 20 players in the GM cohort.}
        \label{fig:perm_gm}
    \end{subfigure}
	\caption{Posterior mean estimates for winner/loser-effects after permuting
the game data. For each the true posterior mean estimate is 
shown in red, with the intervals representing 66\% and 95\% of the 
estimates obtained under the
permutation distributions.}
	\label{fig:gm_perm}
\end{figure}



\paragraph*{Simulation Study}

Our previous analyses reveal that the experiential effects that we are
interested in, if they exist, are likely 
small. 
In this case, even the large number of games analysed here may not
be sufficient to detect such small effects.
To investigate this further we construct a simulation study, examining 
when these effects can be identified from a cohort of 10 simulated focal players.
To do this, we simulate game outcomes from the model proposed in Section~\ref{sec:model}.
The values of the parameters in this model are given by:
\begin{itemize}
    \item $\gamma_1 = 0.091$ 
    \item $\gamma_2 =  0.0038$
    \item $\alpha_j \sim N(-0.075, 0.025)$. This distribution approximates the estimates from Figure~\ref{fig_s4_left}.
    \item $\beta_j \sim N(\mu_\beta, 0.01)$. This variance is close to the posterior mean for $\tau_2$ estimated from the 1700-1900 cohort.
    \item The rating difference for any given game is sampled from a Laplace distribution with mean 0 and scale 30. This closely resembles the actual rating differences in the 1700-1900 cohort.
    \item For any given game, the simulated player had 0.5 probability of being white and 0.5 probability of being black.
\end{itemize}

We choose to only look at the previous game for history here ($n = 1$).
To investigate what effects we can detect we vary the true value of $\mu_{\beta}$
in this simulation along with the number of games played by each player, using 
$\mu_\beta = (-0.5, 0.12, 0.5)$ and $n_\text{per player} = (1,000, \; 5,000, \;20,000)$
We chose two large global experiential effects in different directions and one small 
positive experiential effect, giving a range of possible true effects.
Note that $\mu_\beta = 0.12$ equates to an approximate 
average boost in win percentage by 3\% when coming from a win for all players in the 
dataset. We chose a maximum of 20,000 games per player to be somewhat
larger than the average number of games per player in the 1700-1900 cohort. 
We show the posterior estimates of $\mu_{\beta}$ and $\beta_j$ for each 
of these 9 simulation settings in Figure~\ref{fig_sim}.

 
\begin{figure}[ht]
	\centering
    \begin{subfigure}{0.45\textwidth}
        \centering
        \includegraphics[width=\textwidth]{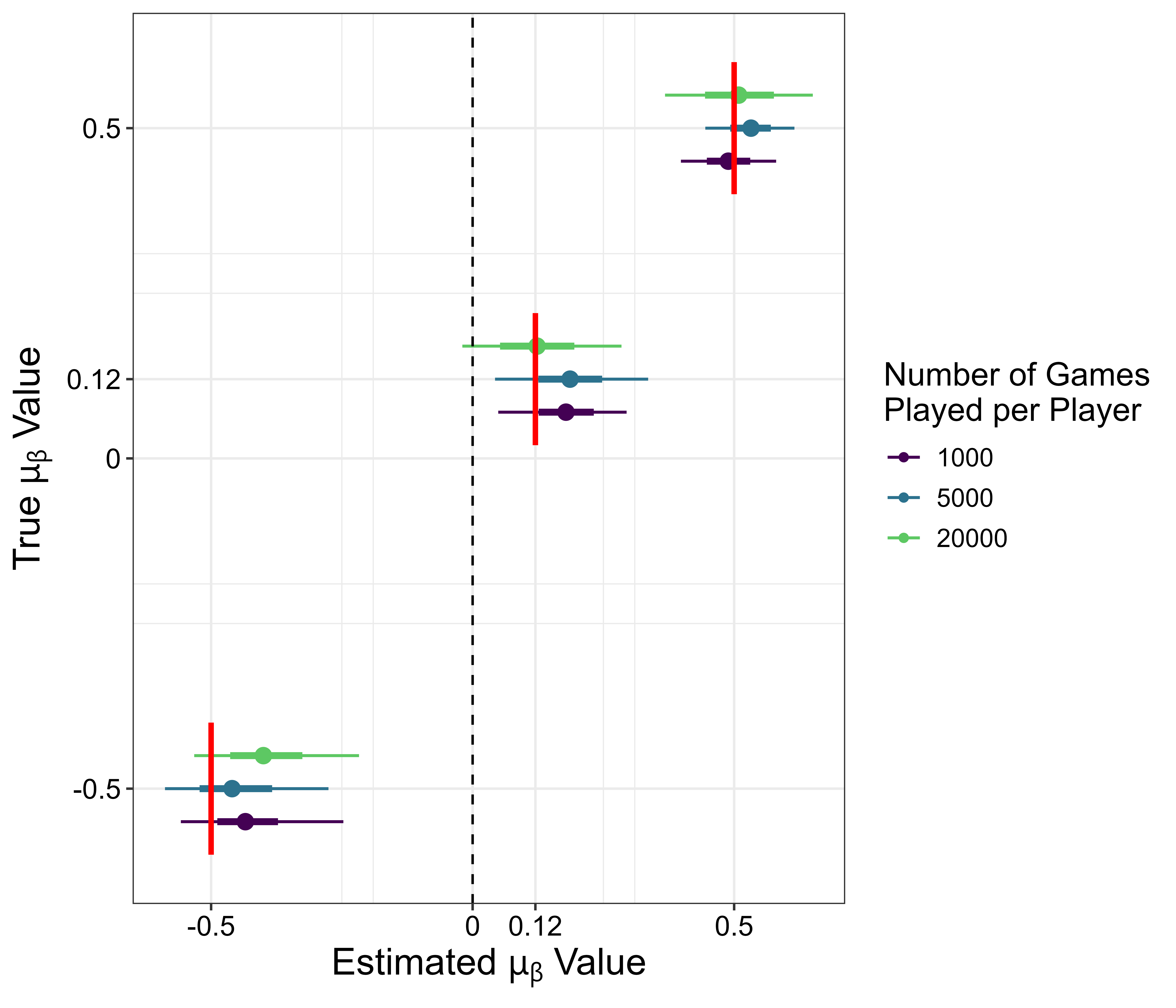}
        \caption{The estimated values of $\mu_\beta$ for 3 different known effect sizes.}
        \label{fig_sim_left}
    \end{subfigure}
    \begin{subfigure}{0.45\textwidth}
        \centering
        \includegraphics[width=\textwidth]{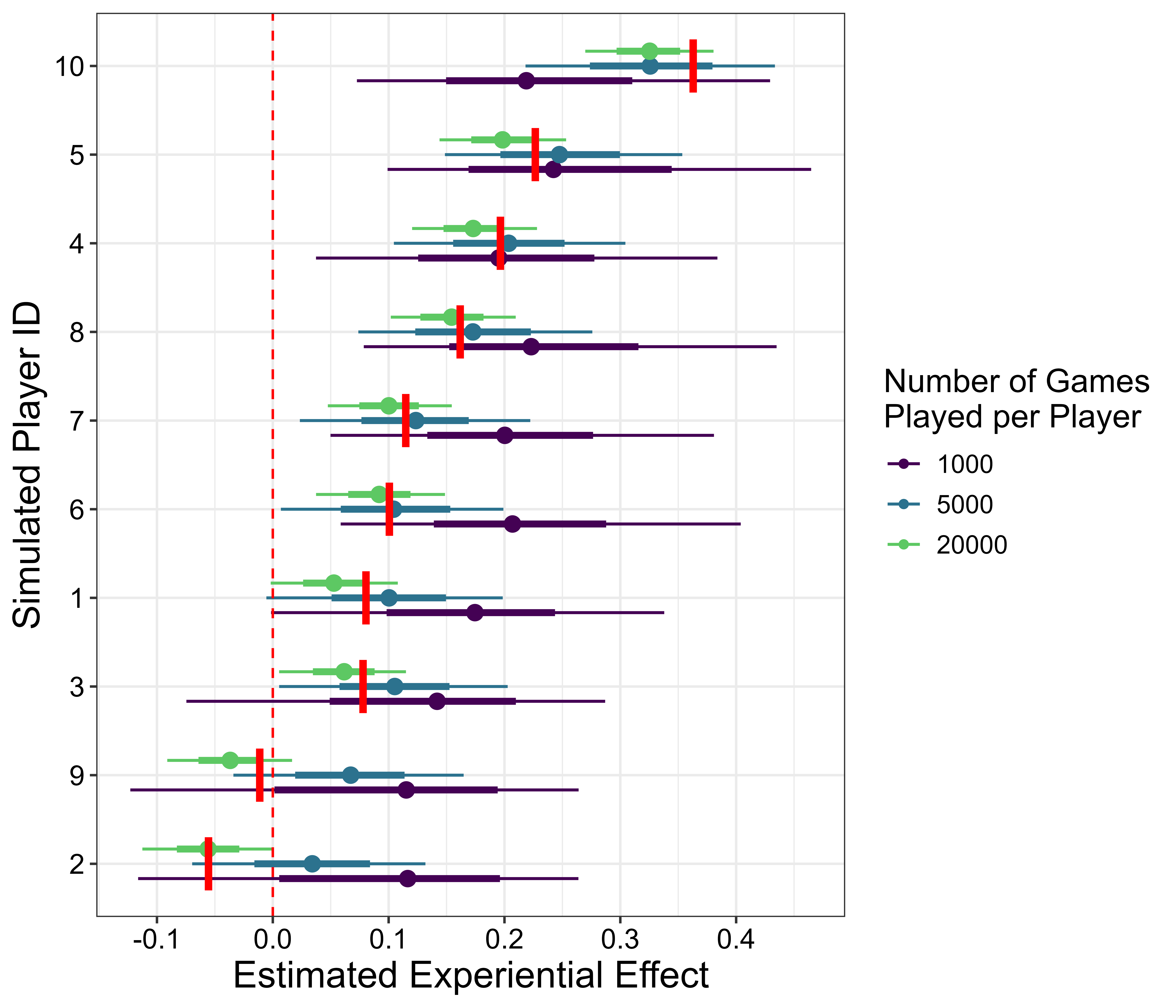}
        \caption{The estimated experiential effects for 10 simulated players for known $\mu_\beta = 0.12$.}
        \label{fig_sim_right}
    \end{subfigure}
	\caption{66\% and 95\% credible intervals for estimated effects for $\mu_\beta$ and $\beta_j$. In both figures the red vertical lines represent the true values of each respective parameter.}
	\label{fig_sim}
\end{figure}

In each simulation study, the 95\% credible intervals contain the true parameter values
for both $\mu_{\beta}$ and each $\beta_j$. We note that when estimating
individual $\beta_j$, we need a large number of games to get posterior intervals
which do not contain 0 when the true value is non-zero. While a large number of games
is required to detect these small effects, as shown in our model checking procedures,
these small effects have little importance in describing the outcomes of these 
games.



\paragraph*{Player Effects}
Finally, we show some of the estimated player effects which we have excluded from the
main text. In Figure~\ref{fig_s4} we show the posterior estimates for each $\alpha_j$
for the 1700-1900 and GM cohort. For the lower ability players many of these effects 
appear to be negative. This agrees with intuition that these players are likely to
win less than half of their games. For the GM cohort we see some strong positive 
player effects, including for Magnus Carlsen. 

\begin{figure}[ht]
	\centering
    \begin{subfigure}{0.45\textwidth}
        \centering
        \includegraphics[width=\textwidth]{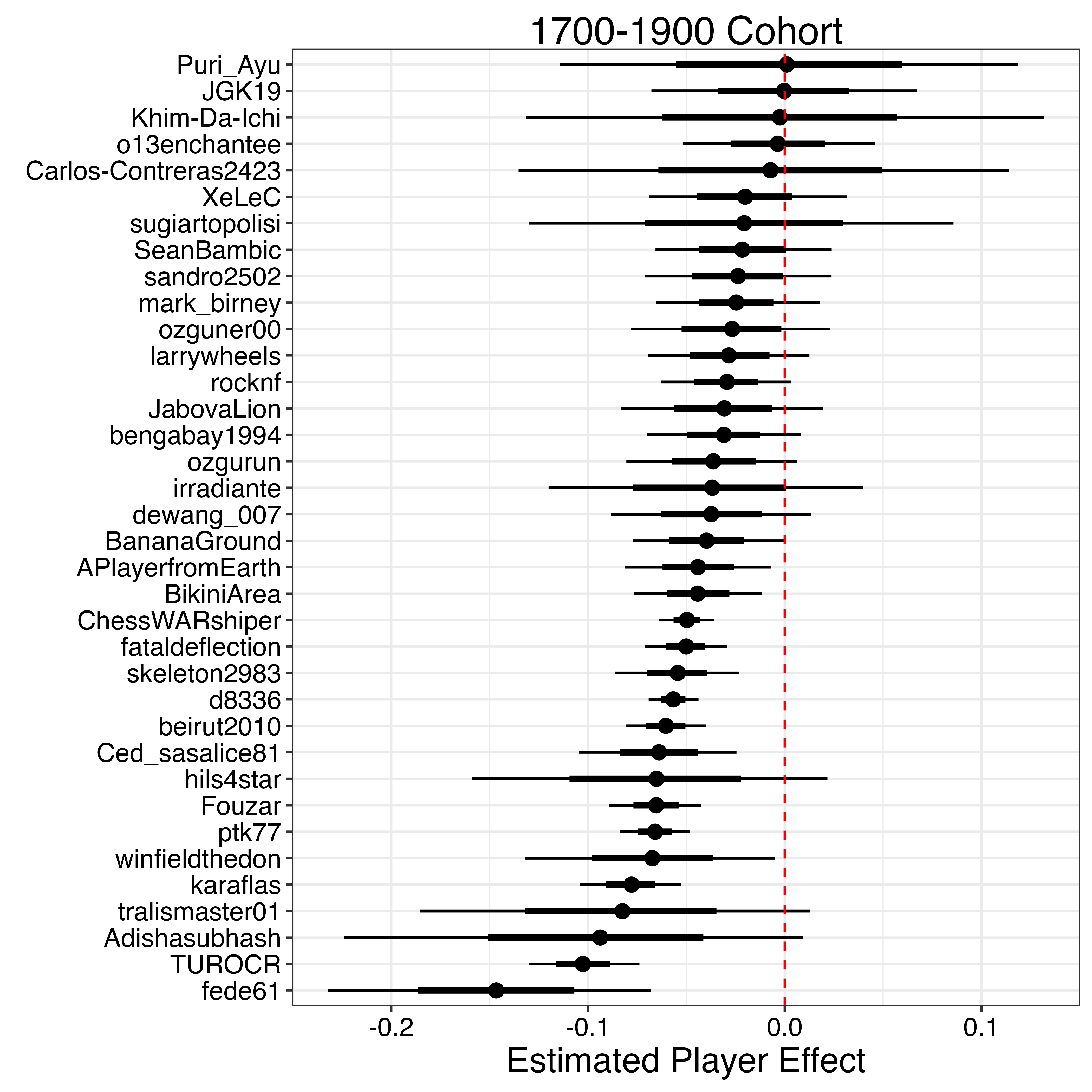}
        \caption{Player Effects for the 1700-1900 cohort.}
        \label{fig_s4_left}
    \end{subfigure}
    \begin{subfigure}{0.45\textwidth}
        \centering
        \includegraphics[width=\textwidth]{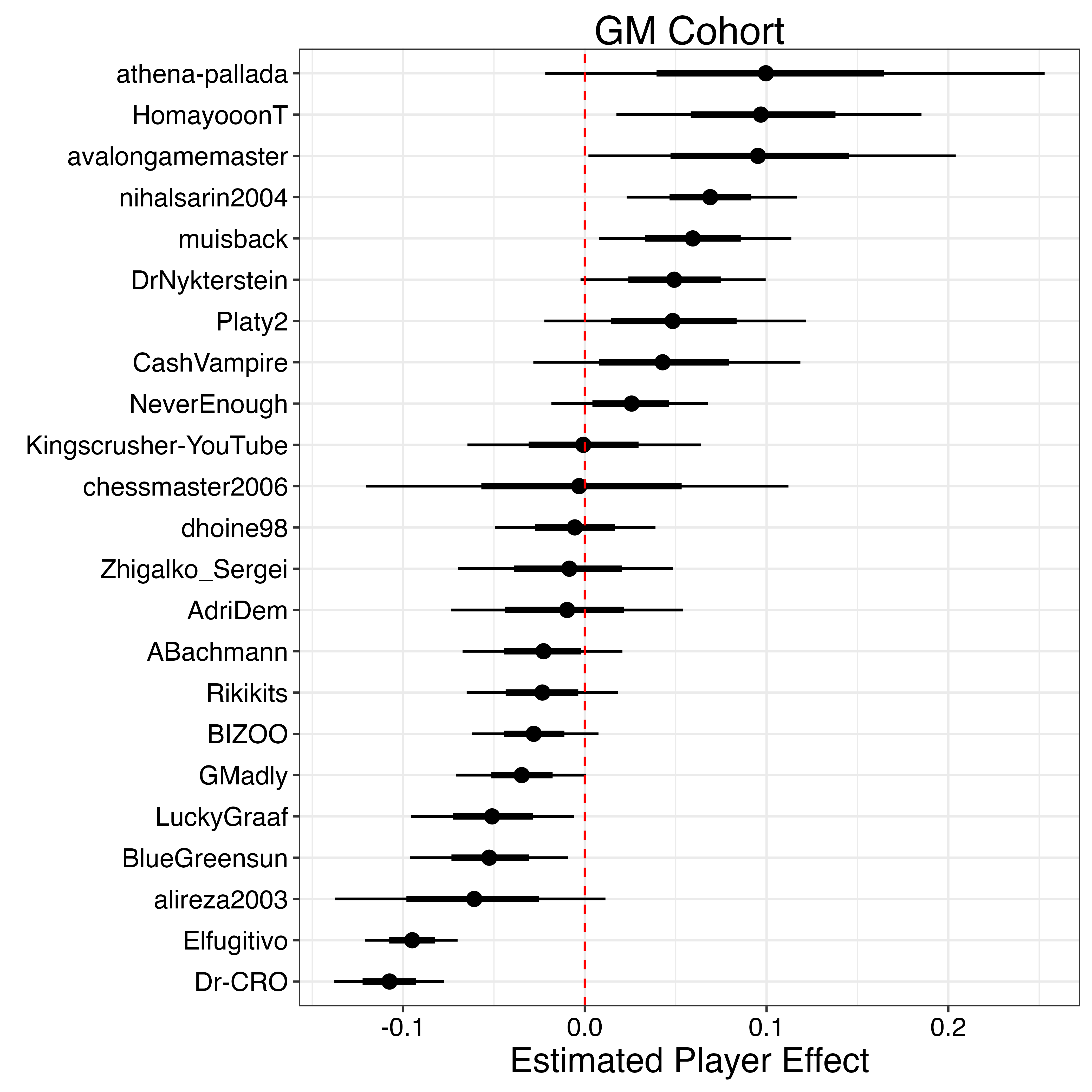}
        \caption{Player Effects for the GM cohort.}
        \label{fig_s4_right}
    \end{subfigure}
	\caption{Estimated Player Effects for the 1700-1900 and GM cohorts.}
	\label{fig_s4}
\end{figure}

\end{appendix}

\printbibliography

\end{document}